\documentclass[fleqn,usenatbib]{mnras}

\usepackage{newtxtext,newtxmath}

\usepackage[T1]{fontenc}

\DeclareRobustCommand{\VAN}[3]{#2}
\let\VANthebibliography\thebibliography
\def\thebibliography{\DeclareRobustCommand{\VAN}[3]{##3}\VANthebibliography}

\usepackage{graphicx}	
\usepackage{amsmath}	
\usepackage{siunitx}
\usepackage{booktabs}
\usepackage[normalem]{ulem}

\newcommand{\Tbar}{\ensuremath{\tilde{T}}}
\newcommand{\rhobar}{\ensuremath{\tilde{\varrho}}}
\newcommand{\ri}{\ensuremath{r_\text{i}}}
\newcommand{\ro}{\ensuremath{r_\text{o}}}
\newcommand{\rmid}{\ensuremath{r_\text{mid}}}
\newcommand{\Msun}{\ensuremath{\mathrm{M_\odot}}}
\newcommand{\vect}[1]{\ensuremath{\mathbf{#1}}}

\title[GW signal of PNS convection]{Gravitational wave signature of proto-neutron star convection:\\ I. MHD numerical simulations}

\author[R. Raynaud et al.]{
Rapha\"{e}l Raynaud,$^{1}$\thanks{E-mail: raphael.raynaud@cea.fr}
Pablo Cerd\'{a}-Dur\'{a}n,$^{2}$
J\'er\^ome Guilet$^{1}$
\\
$^{1}$Universit\'{e} de Paris and Universit\'{e} Paris-Saclay, CEA, CNRS, AIM,
F-91191 Gif-sur-Yvette, France\\
$^{2}$Departamento de Astronom\'ia y Astrof\'isica, Universitat de València, E46100 Burjassot (Valencia), Spain \\
}

\date{Accepted XXX. Received YYY; in original form ZZZ}

\pubyear{2021}

\begin{document}
\label{firstpage}
\pagerange{\pageref{firstpage}--\pageref{lastpage}}
\maketitle

\begin{abstract}
Gravitational waves provide a unique and powerful opportunity to constrain the dynamics in the interior of proto-neutron stars during core collapse supernovae. Convective motions play an important role in generating neutron stars magnetic fields, which could explain magnetar formation in the presence of fast rotation. We compute the gravitational wave emission from proto-neutron star convection and its associated dynamo, by post-processing three-dimensional MHD simulations of a model restricted to the convective zone in the anelastic approximation. We consider two different proto-neutron star structures representative of early times (with a convective layer) and late times (when the star is almost entirely convective). In the slow rotation regime, the gravitational wave emission follows a broad spectrum peaking at about three times the turnover frequency. In this regime, the inclusion of magnetic fields slightly decreases the amplitude without changing the spectrum significantly compared to a non-magnetised simulation. Fast rotation changes both the amplitude and spectrum dramatically. The amplitude is increased by a factor of up to a few thousands. The spectrum is characterized by several peaks associated to inertial modes, whose frequency scales with the rotation frequency. Using simple physical arguments, we derive scalings that reproduce quantitatively several aspects of these numerical results. We also observe an excess of low-frequency gravitational waves, which appears at the transition to a strong field dynamo characterized by a strong axisymmetric toroidal magnetic field. This signature of dynamo action could be used to constrain the dynamo efficiency in a proto-neutron star with future gravitational wave detections.
\end{abstract}

\begin{keywords}
 convection - MHD (magnetohydrodynamics)
 - gravitational waves - stars: magnetars
 - supernova: general - methods: numerical
\end{keywords}



\section{Introduction}

The gravitational wave signal from the collapse of massive stars is an excellent opportunity to learn about the physics of neutron stars. During core collapse supernovae (CCSNe), the core of stars born with masses in the range $M \sim 8-\SI{100}{\Msun}$ collapses\footnote{Stars of more than \SI{10}{\Msun} ultimately develop an iron core, while in the range 8--\SI{10}{\Msun} they develop a O+Ne+Mg core resulting in an electron-capture supernova \citep{Hiramatsu2021}.} and, after reaching nuclear matter density and bouncing, a proto-neutron star (PNS) is formed surrounded by an accretion shock. In a time-scale of 0.1 to \SI{1}{s}, the strong neutrinos flux coming out of the PNS deposits sufficient energy behind the shock to drive a supernova explosion that disrupts the outer layers of the star. This neutrino-driven explosion mechanism is expected to be the dominant driver of supernovae for slowly rotating progenitor cores \citep{Burrows1986,Bethe1990,Janka2001,janka17}. A small fraction of the electromagnetic observation of CCSNe ($\sim 1\%$) show indications of a fast-rotating progenitor (broad-lined type Ic SNe \citep{Li2011} or long duration gamma-ray bursts \citep{Chapman2007}). In those cases the supernova is probably driven by magneto-rotational explosions \citep[see][and references therein]{Bugli2020,Obergaulinger2020}. This scenario has also been suggested as birth place of magnetars, neutron stars with magnetic fields of $10^{16}$~G that could be the result of
dynamo action generated by turbulent convection that develops inside the proto-neutron stars
\citep{Thompson1993,Raynaud2020}.
The magnetar birthrate has been observationally estimated to be at least $10\%$ (with large uncertainties) of the CCNSe rate \citep{Kouveliotou1994,Gill2007,Beniamini2019b}, which would imply that dynamos could be operative in cases in which magneto-rotational explosions are not produced.
Regardless of the mechanism, the explosion leaves behind a compact remnant that in most cases is a hot neutron star. Alternatively the PNS could undergo black-hole formation, although this possibility is observationally constrained to about $15-20\%$ of all CCSNe \citep{Kochanek2014,Adams2017}.

Multidimensional numerical simulations \citep{Murphy2009,Marek2009,Yakunin2010,Scheidegger2010,Muller2012,Muller2013,Yakunin2015,Kuroda2017,Andresen2017,Powell2019, Morozova2018,Mezzacappa2020,Andresen2021} show that the GW signal during neutrino-driven explosions is dominated by the excitation of high frequency (buoyancy driven) $g$-modes in the PNS and low frequency ($\sim\SI{100}{Hz}$) standing shock accretion instabilities \citep[SASI,][]{Kuroda2016} during the period of time elapsed between the bounce and the time of the explosion ($\sim\SI{1}{s}$).
In those cases, the typical rms strain of the GW signal in 3D simulations  is $h\sim 4-15 \times 10^{-24}$ at $10$~kpc, with frequencies ranging from $\sim 50$~Hz to a few kHz \citep{LopezPortilla2020}. For simulations of fast rotating progenitors, the GW strain depends on the rotation rate and can reach values of $\sim 10^{-21}$ at $10$~kpc \citep{Ott2005, Scheidegger2008,Scheidegger2010, Cerda-Duran2013,Kuroda2014,Takiwaki2018,PAjkos2019,Pan2021,Shibagaki2021}. These signals could be observed with current GW detectors, aLIGO \citep{LIGOScientificCollaboration2015}, aVirgo \citep{Acernese2015} and KAGRA \citep{Aso2013}, within $\sim 10$~kpc for neutrino-driven explosions and up to $\sim 100$~kpc for magneto-rotational explosions \citep{Szczepanczyk2021}. Additional details about the GW signal from CCSNe can be found at \cite{kotake17} and \cite{Abdikamalov2020}.

After the onset of the explosion, the amplitude of the two dominant contributions to the GW signal, the excitation of $g$-modes and the SASI, decay very rapidly. The only mass motion in the resulting PNS is the result of the convection driven by the neutrino cooling of the PNS.
Calculations of the cooling of PNSs \citep{Burrows1988,Keil1995,Keil1996,Pons1999,Miralles2000,Roberts2012,Mirizzi2016} show that extended regions of the star can stay convectively active for $3-50$~s after the explosion.
No attempts have been made so far to estimate the GW signature during this phase, the main reason being the difficulty to perform multidimensional simulations of the PNS in such long time-scales, due to the severe time-step restrictions of numerical hydrodynamics codes.
\cite{Ferrari2003,Ferrari2004,Ferrari2007} have studied the appearance of unstable $g$-modes in PNSs as a possible source of GW, concluding that non-linear saturation would likely limit the maximum strain to values unobservable with current detectors. Very recently there has been an increased interest in the study of proto-neutron star convection \citep{Nagakura2020} and its associated dynamo \citep{Raynaud2020,Masada2020}, though these works have been focused on the early post-bounce phase and did not provide a description of the late-time post-explosion phase several seconds after bounce. \cite{Raynaud2020} ran a series of MHD simulations focused on the convective zone, varying in particular the rotation rate. They showed that the efficiency of dynamo action increases for fast rotation rates, in the regime where the Coriolis force is dominant compared to the non-linear advection term (i.e. at low Rossby numbers). In this regime, they obtain a new kind of strong field dynamo solution with a magnetic field dominated by its axisymmetric azimuthal component. The magnetic energy is up to ten times larger than the kinetic energy and follows a magnetostrophic scaling, in which the Lorentz force balances the Coriolis force. This dynamo solution is obtained above a critical rotation rate and \citet{Raynaud2020} have shown that it has the potential to explain the formation of magnetars.

We aim at estimating for the first time the characteristic strain, frequency and spectral features present in the post-explosion GW emission of PNSs within the first
\SI{10}{s} after bounce. In this signal, we also look for a signature of the magnetic fields generated by dynamo action, as this could provide a unique testbed for PNS dynamo models and magnetar formation theories. We base our calculations on the 3D simulations of the convective zone of PNSs of \cite{Raynaud2020} and an extension of this model with a PNS structure representative of late times several seconds after bounce (Raynaud \emph{et al.}, in prep).

This work is organized as follows: in Section~\ref{sec:modelling} we describe the numerical setup used in the numerical simulations, in Section~\ref{sec:results} we present the results of the GW signals in the different regimes studied and find scaling relations for the amplitude and typical frequencies in terms of the properties of the PNS. We present our conclusions in Section~\ref{sec:conclusions}.

\section{Modelling}\label{sec:modelling}

Our study directly follows from the new setup introduced by \citet{Raynaud2020} to study PNS convective dynamo as a scenario of magnetar formation. While recalling the main features and the underlying hypotheses of this setup, we refer the reader to the above article for further details. The gravitational wave signal generated by proto-neutron star convection is computed from three-dimensional (3D) numerical
simulations solving non-linear, magneto-hydrodynamic equations
governing the flow of an electrically conducting fluid in a rotating
spherical shell. Energy-based scalings are consistent with convective
velocities of order $\SI{e8}{cm/s}$ \citep{Thompson1993}, and hence a
relatively low Mach number turbulence ($M\sim 10^{-2}$), since the
typical sound speed in the convective zone ranges from $0.2 c $ to
$0.6 c$ ($c$ being the speed of light).
This justifies the use of the soundproof anelastic approximation to describe the convective fluid motions, which are treated as perturbations from an isentropic background state assumed at mechanical (hydrostatic) and thermal (isentropic) equilibrium \citep{Jones2011}.

A major difficulty in modelling neutron star formation lies in the
treatment of neutrinos. In our case however, this can also be greatly
simplified assuming that the neutrinos are in thermodynamic
equilibrium with matter within the hot and dense newborn PNS.  At
length scales much larger than the neutrino mean free path ($\lambda
\sim \SI{10}{m}$), energy and momentum transport can be approximated
by the diffusion approximation. This approximation is, of course, less
and less justified as we get closer to the PNS surface. It is nonetheless
a fair hypothesis in our case, since we will focus on the convective zone that
lies well inside the neutrinosphere.

With our approach, the complexity of neutrino-matter interactions is
then reduced to the transport coefficients controlling viscous and
thermal diffusion, while all the PNS characteristics are embedded
within the prescription of the adiabatic background, which not only
includes the density and temperature stratification of the convective
zone but also the thermodynamics coefficients entering the physics of
thermal convection. In what follows, all our knowledge of the PNS
structure and evolution encapsulated in the adiabatic background is
extracted from a 1D core-collapse supernova simulation that describes
the formation of neutron star with a baryonic mass of \SI{1.78}{\Msun}
\citep{Hudepohl2010,Hudepohl2014}. The calculations were performed with the code
\textsc{Prometheus-Vertex} \citep{Rampp2002} and use the non-rotating
\SI{27}{\Msun} progenitor s27.0 by Woosley et al. \citep{Woosley2002}
and the equation of state LS220 \citep{Lattimer1991}.

\begin{figure}
  \includegraphics[width=\columnwidth]{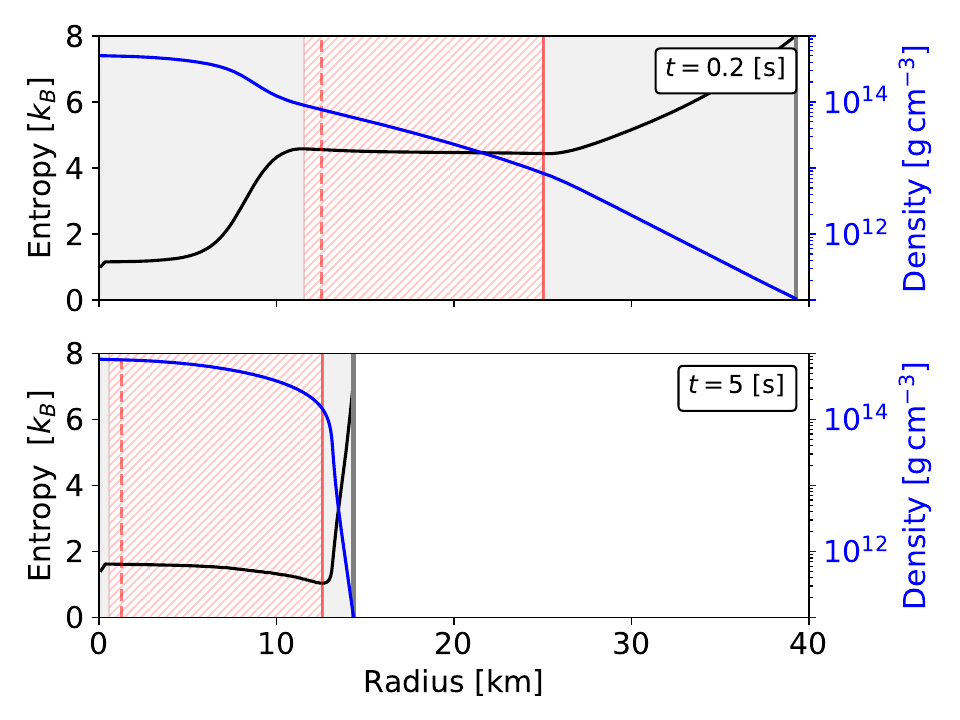}
  \caption{Evolution of the proto-neutron star structure. The black
    (blue) line shows the entropy per baryon (density) as a function
    of radius. The red hatched area shows the extent of the convective
    zone according to the Schwarzschild criterion $\partial S/\partial
    r<0$ and the grey areas indicate stably stratified zones. The
    vertical grey line indicates a proxy for the PNS surface defined
    by $\tilde{\varrho}(R)=\SI{e11}{g.cm^{-3}}$. The solid red line
    indicates the outer radius of the convective zone~\ro{} while the
    dotted line indicates the inner radius~\ri{} delimiting the 3D
    computational domain.}
  \label{f:back}
\end{figure}

Last but not least, we will make a strong approximation concerning the
time evolution of the PNS structure (see Fig.~\ref{f:back}).
Indeed, the simultaneous computation of the turbulent convection and
the comparatively slow contraction of the PNS from $R\sim \SI{39}{km}$
at $t=\SI{0.2}{s}$ post-bounce to $R\sim\SI{14}{km}$ at $t=\SI{5}{s}$
would require to include overlying and underlying stably stratified
layers in the computational domain and a time-evolving background,
which is far beyond the scope of the present study that aims at a
first characterization of the GW signal generated by rotating
convection.  We thus restrict the simulation domain to a spherical
shell corresponding to the convective region. We also keep the
anelastic background steady and consider instead two independent sets
of numerical models with two different adiabatic backgrounds
representative of the PNS convective zone at ``early'' time
($t=\SI{0.2}{s}$) and ``late'' time ($t=\SI{5}{s}$), respectively.
Table~\ref{tab:A} summarizes the important characteristics of the
convective zone derived from the 1D core-collapse model. We see in
Fig.~\ref{f:back} that we did not try to match exactly the size of the
convective zone predicted by the 1D model (solid and dashed red
line). At $t=\SI{0.2}{s}$, we approximate the shell aspect ratio to
$\chi=\ri/\ro=0.5$. At $t=\SI{5}{s}$, we artificially increase the
shell aspect ratio to avoid technical difficulties that may arise with
a setup approaching a full sphere.
\begin{table}
\centering
  \caption{Characteristics of the convective zone. The subscript
    $_\text{o}$ designates quantities evaluated at the outer edge of
    the convective zone at $r=\ro$. The sound speed $c_s$, adiabatic
    index $\gamma$ and the temperature gradient are evaluated in the
    middle of the convective zone.
    }\label{tab:A}
  \begin{tabular}{lcc}
    \toprule
    time post-bounce [s]  & 0.2 &  5\\\midrule
    \ro{}  [km]  & 25 & 12.6 \\
    $r_\text{mid}$  [km]  & 18.3 & 6.59 \\
    $\tilde{\varrho}_\mathrm{o}$  [\si{g.cm^{-3}}] & $8.33 \times 10^{12} $ & $1.45 \times 10^{14}$ \\
    $\tilde{\varrho}_\mathrm{mid}$  [\si{g.cm^{-3}}] & $3.06 \times 10^{13} $ & $6.11 \times 10^{14}$ \\
    $\tilde{T}_\mathrm{o} $  [\si{K}] & $ 1.36 \times 10^{11} $  & $ 8.72 \times 10^{10} $ \\
    $\Phi_\mathrm{o}$ [\si{erg.s^{-1}}] & $2.16 \times 10^{52}$ & $1.13 \times 10^{51}$ \\
    $M_\text{conv}/\Msun$    & 0.83 & 1.70 \\
    ${c_s}/c$ & 0.198 & 0.555\\
    $\gamma$ & 1.43 & 2.14\\
    $\left|\partial_r \ln \tilde{T}\right|$ [\si{cm^{-1}}] & $9.43\times10^{-7}$ & $8.74 \times 10^{-7}$ \\
    \bottomrule
  \end{tabular}
\end{table}

This way, we replace the continuous evolution of the PNS by two
``snapshots'' representative of a proto-neutron star at two different
times.  Although this will make the interpretation of the GW waveforms
more difficult, our simplified models allow us to capture the full
complexity of magneto-hydrodynamic processes and the dynamo
amplification of the magnetic field, which makes them complementary to
global core-collapse supernova models.

We carried out a systematic parameter study consisting of 20
early-time and 36 late-time models. With both types of models, we
mainly vary the PNS rotation rate (through the Ekman number -- see
Sect.~\ref{s:eq}) to probe different dynamical regimes.
In hydrodynamic models, the velocity field is initialised with a solid-body rotation. For 8 models, as in \citet{Raynaud2020}, we added to the saturated turbulent state of the hydrodynamic simulation a seed magnetic field whose amplitude is small enough to clearly observe the kinematic growth of the dynamo. However, to avoid the systematic computation of the transient kinematic growth which is computationally expensive, we often restarted from a ``nearby'' solution in the parameter space to ensure a faster convergence toward the equilibrium solution of the new parameter set (of course, this procedure stands only for models sharing the same anelastic background). The initial value of the magnetic field is summarized in Table~\ref{tab:models} in the Appendix.
Our database includes rotation periods ranging from milliseconds to seconds, which allows us to cover about three orders of magnitude in terms of Rossby number, which is a common dimensionless measure of the importance of the non-linear advection term with respect to the Coriolis force.

\subsection{Governing equations}\label{s:eq}

Under the above assumptions, the dynamics in the convective zone of
the PNS is governed by the Lantz-Braginsky-Roberts anelastic equations
\citep{Braginsky1995,Lantz1999}. In the reference frame of the PNS
rotating at angular frequency $\Omega$, they read
\begin{equation}
  {\nabla} \cdot \left(\tilde{\varrho}\textbf{u}\right) = 0 \,,
  \label{e:mass}
\end{equation}
\begin{multline}
  \frac{D \textbf{u}}{D t}
  = -\frac{1}{E} {\nabla} \left(\frac{p}{\tilde{\varrho}}\right) -
  \frac{2}{E} \textbf{e}_z\times \textbf{u} - \frac{Ra}{Pr} \frac{d
    \tilde{T}}{dr} S \textbf{e}_r  \\ + \frac{1}{E Pm}\frac{1}{\tilde{\varrho}} \left(
       {\nabla}\times \textbf{B} \right) \times \textbf{B} + \textbf{F}^\nu \,,
       \label{e:NS}
\end{multline}
\begin{equation}
  \tilde{\varrho}\tilde{T} \frac{D S}{D t} =
  \frac{1}{Pr} \left({\nabla}\cdot\left(
  \tilde{\kappa} \tilde{\varrho}\tilde{T} {\nabla} S \right) +  H \right)
  + \frac{Pr}{Ra}
  \left( Q^\nu + Q^j \right)
  \,,
  \label{e:S}
\end{equation}
\begin{equation}
  \frac{\partial \textbf{B}}{\partial t} = {\nabla} \times
  \left(\textbf{u}\times \textbf{B}\right)
  - \frac{1}{Pm} {\nabla} \times \left(\tilde{\eta}{\nabla}\times\textbf{B} \right)
  \,,
  \label{e:B}
\end{equation}
\begin{equation}
  {\nabla} \cdot \textbf{B} = 0 \,.
  \label{e:divB}
\end{equation}
To make these equations dimensionless, we use the shell gap
$d=\ro-\ri=\ro\left(1-\chi\right)$ as reference length scale, where
$\ro$ and $\ri$ are the inner and outer radius of the convective zone
and $\chi=\ri/\ro$. We fix $\chi=0.5$ for the models at
$t=\SI{0.2}{s}$ and $\chi=0.1$ for the models at
$t=\SI{5}{s}$. The viscous time ${d^2}/{\nu_\mathrm{o}}$ serves as
reference time scale.  We write $\nu_\mathrm{o}$, $\kappa_\mathrm{o}$
and $\eta_\mathrm{o}$ the values at $r=\ro$ of the kinematic
viscosity, thermal diffusivity and magnetic diffusivity and use them
to rescale the corresponding profiles $\tilde{\nu}(r)$,
$\tilde{\kappa}(r)$ and $\tilde{\eta}(r)$. Similarly, we scale the
background density~$\tilde{\varrho}(r) $ and
temperature~$\tilde{T}(r)$ by their values at the outer edge,
$\tilde{\varrho}_\text{o}$ and $\tilde{T}_\text{o}$. The velocity~$\vect{u}$ is
expressed in units of $\nu_\text{o}$/$d$, the magnetic field~$\vect{B}$ in units of
$\sqrt{\Omega \tilde{\varrho}_\mathrm{o} \eta_\mathrm{o} \mu_0}$ where
$\mu_0$ is the vacuum permeability, and the entropy $S$ in units of $\left|d \partial S / \partial r \right|_{\ro}$.

We used only entropy to describe the buoyancy force because it is the dominant driver of convection in the 1D models on which we base our setup. The size of the convective zone is indeed not affected by the inclusion of the lepton number gradient in the stability criterion. The lepton number gradient may, however, play an important role according to other models \citep[e.g.][]{Nagakura2020}. In the limit where the thermal and lepton number diffusivities are assumed to be identical, the anelastic equations can be written such that the entropy~$S$ can be seen as a codensity that describes buoyancy effects associated with both entropy and lepton number gradients \citep[see][Sect.~4.2]{Braginsky1995}. We therefore expect our results to hold at least qualitatively also in cases where lepton-number gradient is important.

The viscous force $\textbf{F}^\nu$ and the viscous heating~$Q^\nu$ are
defined by $F^\nu_i= \tilde{\varrho}^{-1} \partial_j \sigma_{ij}$ and
$Q^\nu = \partial_j u_i \sigma_{ij}$, where the rate of strain tensor
$\sigma_{ij}= 2\tilde{\varrho} \tilde{\nu} \left(e_{ij} -
e_{kk}\delta_{ij}/3 \right)$ and the deformation tensor $e_{ij} =
(\partial_j u_i + \partial_i u_j)/2$ are expressed with the Einstein
summation convention and the Kronecker delta~$\delta_{ij}$.  The Joule
heating entering the energy equation is defined by $Q^j={\tilde{\eta}}
\left({\nabla} \times \textbf{B}\right)^2/{({Pm^2E})}$.

Convection is driven by a fixed energy flux at the outer boundary
$\Phi_\mathrm{o}=4 \pi r_\mathrm{o}^2 \kappa_\mathrm{o}
\tilde{\varrho}_\mathrm{o} \tilde{T}_\mathrm{o} \left.{\partial
  S}/{\partial r}\right|_{r_\mathrm{o}}$.  For the models at
$t=\SI{0.2}{s}$ the energy flux at the inner boundary is fixed so that it
balances the outer heat flux and the internal heat source term $H$ is
set to zero. For the models at $t=\SI{5}{s}$ that are characterized by
a smaller aspect ratio $\chi=0.1$ (larger shell), we impose a zero
flux condition at the inner boundary and we introduce a homogeneous
internal heat source term $H$ to balance the outer energy flux \citep{Lepot2018}.
For the models at $t=\SI{0.2}{s}$, we impose perfect conductor magnetic
boundary conditions, i.e.  $\vect{B}\cdot\vect{n}=0$ with $\vect{n}$
the unit vector normal to the boundary. For the models at
$t=\SI{5}{s}$, we keep the perfect conductor condition at the inner
boundary and use perfect conductor or pseudo-vacuum
($\vect{B}\times\vect{n}=0$) condition at the outer boundary.
Finally, we apply stress-free boundary conditions for the velocity
field, $u_r=\partial \frac{u_\theta}{r} / \partial r = \partial
\frac{u_\phi}{r} / \partial r =0$.

The remaining physical control parameters entering the Navier-Stokes,
energy and induction equations are the Ekman, Rayleigh, thermal and
magnetic Prandtl numbers defined respectively by
\begin{equation}
  E = \frac{\nu_o}{\Omega d^2} \,,\quad
  Ra= \frac{\tilde{T}_\mathrm{o} d^3 \left.\frac{\partial S}{\partial r}\right|_{r_\mathrm{o}}}{\nu_\mathrm{o} \kappa_\mathrm{o}} \,,\quad
  Pr = \frac{\nu_\mathrm{o}}{\kappa_\mathrm{o}} \,,\quad
   Pm = \frac{\nu_\mathrm{o}}{\eta_\mathrm{o}}
   \,.\label{e:param}
\end{equation}
We proceed as in \citet{Raynaud2020} to express the simulations
outputs in physical units. We first deduce the PNS rotation rate with
the relation
\begin{equation}
\Omega = \left(\frac{\Phi_\mathrm{o}}{4 \pi r_\mathrm{o}^2
    \tilde{\varrho}_\mathrm{o}  d^3} \frac{Pr^2}{E^3 Ra}\right)^{1/3}\,.
\end{equation}
The values of the diffusivities follow from the definitions of the Ekman and Prandtl numbers in Eq.~\eqref{e:param}.
Thus, our rescaling process relies on the typical values of 
the convective zone outer radius $r_\mathrm{o}$, density  $\tilde{\varrho}_\mathrm{o}$ and width $d$ on the one hand, and the energy flux~$\Phi_\mathrm{o}$ on the other hand. One may wonder to what extent all these values are sensitive to the lepton fraction gradient that is also expected to be driving convection in a real proto-neutron star. In this regard, a first point to note is that the values we use (see Table~\ref{tab:A}) are independent of the simplifications we make to set up the 3D model, since we used a 1D core-collapse model to estimate these quantities. Further, we stress that this 1D model takes into account the destabilizing effect of both the entropy and lepton fraction gradients, which enabled us to check that we obtain exactly the same results when using the Schwarzschild or Ledoux criterion to determine the localisation of the convective zone. An uncertainty remains regarding the energy flux $\Phi_{\rm o}$ that we estimate as half of the total neutrino luminosity at the outer edge of the convective zone and whose time evolution is shown in Fig.~\ref{f:phi_o}. We find that our estimate agrees within \SI{30}{\%} with the turbulent energy flux obtained with more advanced 3D core-collapse supernova models, at least at $t=\SI{0.2}{s}$ \citep[][Fig.~16]{Nagakura2020}. Last but not least, even if the numerical values of the characteristic parameters of the GW signal might change with finer estimates of the above quantities, we stress that (i) these must currently be taken with care since we do not take into account the background evolution, (ii) we believe that the 1D model provides us with at least good order of magnitudes to estimate the quantities we need and (iii) this will not affect the physical analysis and the scaling relations we will derive.

\subsection{Background implementation}

\begin{figure}
\includegraphics[width=\columnwidth]{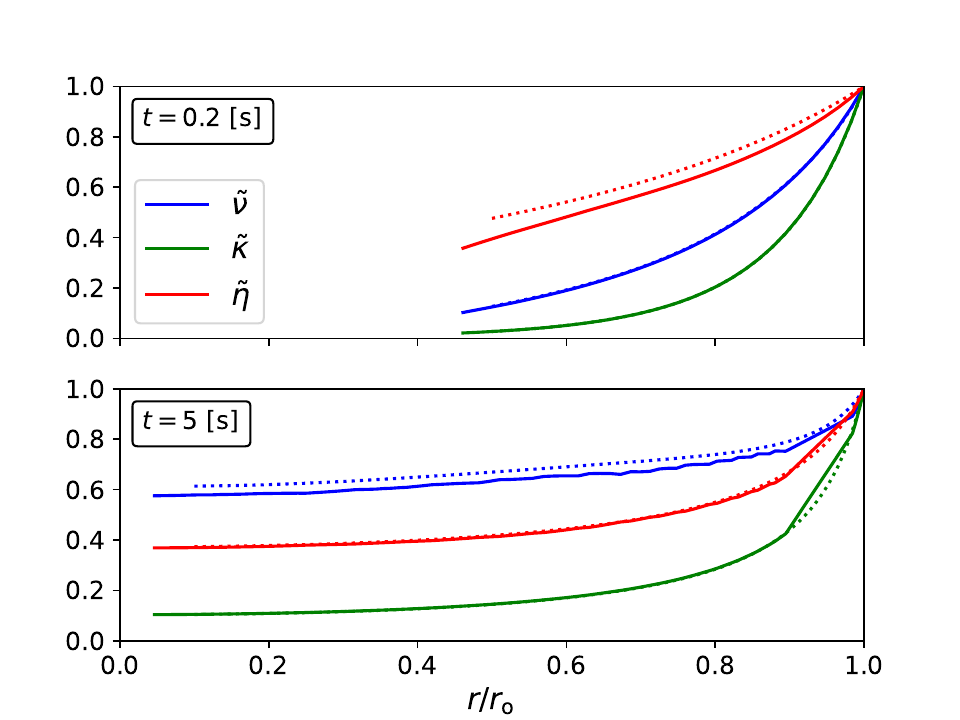}
    \caption{Normalized diffusivity profile as a function of
      radius. The solid lines show the profiles
      from the 1D simulation
      and the dotted lines show the profiles implemented in the code
      (approximated by a polynomial fit or a scaling relation as explained in the text).}
    \label{f:transport}
\end{figure}

The background density $\tilde{\varrho}$ and temperature $\tilde{T}$
profiles at $t=\SI{0.2}{s}$ and $t=\SI{5}{s}$ post-bounce are
implemented with fifth order polynomials that fit the outputs of the
1D core-collapse supernova model.  Fig.~\ref{f:transport} displays the
radial profiles of the viscous, thermal and magnetic diffusivities
overplotted with their approximations implemented in the code.
For the neutrino kinematic viscosity, we use the approximation
\citep{Guilet2015b}
\begin{equation}
  \tilde{\nu} \propto \tilde{T}^2 \tilde{\varrho}^{-2}
  \,.
  \label{e:nu}
\end{equation}
Assuming each neutrino opacity is proportional to the square of the
neutrino energy $\chi_\nu = {\chi_0} (h \nu)^2/{E_0^2}$
\citep{Socrates2005}, the neutrino thermal diffusivity is given by
\citep{Raynaud2020}
\begin{equation}
  \kappa = \frac{1}{36}\frac{1}{\tilde{\varrho} c_p}\frac{k_\mathrm{B}^2
    T}{c^2\hbar^3} \left[ \left(\frac{E_0^2}{\chi_0}\right)_{\nu_e} +
    \left(\frac{E_0^2}{\chi_0}\right)_{\bar{\nu}_e} + 4
    \left(\frac{E_0^2}{\chi_0}\right)_{\nu_x} \right] \,,
  \label{e:thermalcond}
\end{equation}
where the $x$ subscript refers to muon and tau neutrino and
antineutrino flavours, $k_\text{B}$ the Boltzmann constant and $\hbar$
the reduced Planck constant. At $t=\SI{0.2}{s}$, we perform a
polynomial fit of Eq.~\eqref{e:thermalcond} in which we use the
opacities given by \cite{Janka2001a}. At $t=\SI{5}{s}$, the radial
profile of the thermal diffusivity is well approximated by the scaling
\begin{equation}
  \tilde{\kappa} \propto \tilde{\varrho}^{-4/3}
  \,,
\end{equation}
which derives from Eq.~\eqref{e:nu} and the Prandtl number scaling $Pr
\propto \tilde{T}^2\tilde{\varrho}^{-2/3}$ \citep{Thompson1993}.

Finally, the magnetic diffusivity $\eta$ of degenerate, relativistic
electrons scattering on non-degenerate protons scales as
\citep{Raynaud2020}
\begin{equation}
  \eta \propto \left(\tilde{\varrho} Y_e\right)^{-1/3}
  \,,
\end{equation}
where $Y_e$ is the electron fraction. At $t=\SI{0.2}{s}$, we can
neglect the $Y_e$ dependency (dotted vs solid red lines in the top
panel of Fig.~\ref{f:transport}), whereas we fit the profile at
$t=\SI{5}{s}$. When looking at Fig.~\ref{f:transport}, note that the
difference of the radial extent between the top and bottom curves
reflects the evolution of the spherical shell geometry approaching a
full sphere at $t=\SI{5}{s}$. The difference between solid and dashed
lines shows again that we did not try to match exactly the shell aspect
ratio predicted by the 1D model (for the reasons mentioned above).

\subsection{Numerical methods}

The system of Eqs.~\eqref{e:mass}-\eqref{e:divB}, completed by the set
of boundary conditions indicated above, is integrated in time with the
pseudo-spectral code MagIC \citep{Gastine2012}. The code uses
poloidal-toroidal decomposition to satisfy the solenoidal constraints
\eqref{e:mass} and \eqref{e:divB}. The poloidal and toroidal
potentials describing the velocity and magnetic fields and the entropy
and pressure fields are expanded on a basis of spherical harmonics in
the angular directions and of Chebyshev polynomials in the radial
directions. The code relies on the SHTns library to compute the
spherical transforms \citep{Schaeffer2013}.

The time integration is performed with mixed implicit/explicit
algorithms (IMEX), in which are treated explicitly all the terms
coupling different harmonic modes (namely the non-linear terms and the
Coriolis force) . The default scheme is a combination of
Crank-Nicolson /Adams-Bashforth second order schemes
\citep{Glatzmaier1984}.  We also make use of IMEX Runge-Kutta
multistage methods implemented in MagIC, of second
\citep[PC2,][]{jameson81} and third order
\citep[BPR353,][]{Boscarino2013}. We refer the reader to the online
code documentation\footnote{\url{https://magic-sph.github.io}} for a more comprehensive presentation
of the numerical methods \citep[see also][and references
  therein]{Gastine2020}.

\subsection{Gravitational wave computation}

In the slow-motion approximation ($v \ll c$, $GM/Rc^2 \ll 1$), the
gravitational wave emission from a source with compact support can be
computed using the quadrupole formula \citep{Landau75,Misner73}:
\begin{equation}
    h^{\rm TT}_{ij}({\bf X}, T) = \frac{2}{D}\frac{G}{c^4} P_{ijkl} \, \ddot Q_{kl}(t),
    \label{e:quadrupoleFormula}
\end{equation}
where ${\bf X}$ is the vector position of the source's center,
$D=|\bf{X}|$ is the distance to the source, $T=t+D/c$ is the advanced time,
$P_{ijkl}$ is the transverse-traceless projector operator \citep[see section 36.10 in ][]{Misner73},
we use dots to denote time derivatives and $Q_{ij}$ is its reduced mass quadrupole moment defined as
\begin{equation}
    Q_{ij}(t) \equiv \int \varrho({\bf x}, t) \left ( x^i x^j - \frac{1}{3} \delta^{ij} r^2\right) d^3{\bf x}.
\end{equation}
The two polarizations of the gravitational signal can be computed \citep{Oohara1997} as $h_+ = (h^{\rm TT}_{\theta\theta} - h^{\rm TT}_{\varphi\varphi})/2$ and $h_\times = h^{\rm TT}_{\theta\varphi}$, where we have expressed $h^{\rm TT}_{ij}$ in the orthonormal basis associated to the spherical coordinates $(r, \theta, \varphi)$. It is possible to express the GW strain in single complex scalar $h=h_+- i h_\times$.

For spheroidal objects it is convenient to express the strain in terms of the mass multipole moments
(see Appendix \ref{sec:quadrupole} for details):
\begin{align}
h ({\bf X}, T)& =  \frac{1}{D} \frac{G}{c^4}\frac{8\pi}{5} \sqrt{\frac{2}{3}} \sum_{m=-2}^{+2}  \ddot Q_{2m}(t) \,\, {}^{}_{-2}Y^{2m}(\Theta,\Phi), \label{eq:hsph}
\end{align}
where ${}_sY^{lm}$ are the spin-weighted spherical harmonics; $\Theta$ and $\Phi$ are the angles associated to the ${\bf X}$ in spherical coordinates and are directly related to the location of the source in the sky; finally
\begin{equation}
    Q_{lm} \equiv  \int \varrho ({\bf x}, t) r^2 Y_{lm}^\star(\theta,\varphi)  d^3{\bf x}
    \label{e:Qlm}
\end{equation}
are the mass multipole moments, where $Y_{lm}$ are the spherical harmonics.

In the anelastic approximation used in our numerical simulations, the
density is formally equal to its background value $\tilde\varrho$ and
assumed steady in time.  However, it is still possible to evaluate the
density fluctuations $\varrho' = \varrho - \tilde\varrho$, which are
the only ones contributing to $\ddot Q_{lm}$.  The difficulty here is
that the density fluctuations are not part of the time-stepped
variables by the code since they do not appear explicitly in the
anelastic equations. We have to evaluate the density fluctuations as a
function of the entropy~$S$ and pressure perturbations~$p$, according
to the thermodynamics relation
\begin{equation}
  \varrho' = \frac{\tilde{\alpha}}{c_p} (- \Tbar \rhobar S + \frac{1}{\Gamma} p)
  \label{e:rhoprime}
\end{equation}
where the Gr\"{u}nesein parameter~$\Gamma$, the thermal
expansion~$\tilde{\alpha}$ and the specific heat at constant
pressure~$c_p$ are defined by
\begin{align}
 \Gamma& = \left(\frac{\partial \ln \tilde{T}}{\partial \ln
   \tilde{\varrho}}\right)_S \,,\\
 \tilde{\alpha} &=-\frac{1}{\tilde{\varrho}}\left(\frac{\partial \tilde{\varrho}}{\partial \tilde{P}}\right)_P\,,\\
 c_p &= \tilde{T} \left(\frac{\partial S}{\partial \tilde{T}}\right)_P\,,
\end{align}
respectively. In practice, the ratio $\tilde{\alpha}/c_p$ is related
to the adiabatic temperature gradient $\lvert \partial_r\ln \tilde{T}
\rvert= \tilde{\alpha} g / c_p$ and the background density,
temperature and gravity are fitted from our 1D supernova model.

To evaluate the time varying part of $Q_{lm}$, we make use of the
spectral representation used in the code MagIC,
which, e.g,  for $\varrho$ is
\begin{equation}
    \varrho'_{lm}(r)
= \int_0^\pi\int_{0}^{2\pi} \varrho'({\bf x}, t) Y_{lm}^\star(\theta,\varphi) \sin\theta d\theta d\phi.
\end{equation}
Using the spectral variables, we can write
\begin{equation}
  Q'_{lm}
  = \int r^2 \varrho'  Y_{lm}^\star \, d^3 {\bf x}
  =  \int_{\ri}^{\ro} dr \, r^4 \varrho'_{lm}(r)\,, 
\end{equation}
such that $\ddot Q_{lm} = \ddot Q'_{lm}$. $\rho'_{lm}$ is computed
using Eq.~\eqref{e:rhoprime} with the spectral decomposition of the
entropy, $S_{lm}$, and of the pressure, $p_{lm}$.  Finally, to compute
the second time derivatives, we must take into account the change of
reference frame,
\begin{equation}
  \left.\frac{d^2}{dt^2}\right|_\text{obs} X = \frac{d^2}{dt^2} X +
  \frac{d}{dt} \left(\frac{2}{E}\frac{d}{d\varphi} X \right) +
  \frac{1}{E^2} \frac{d^2}{d\varphi^2} X
  \,.
  \label{e:dt2}
\end{equation}
To evaluate this expression, the code outputs the time series
$\left\{t, X, \frac{2}{E}\frac{d}{d\varphi}X, \frac{1}{E^2}
\frac{d^2}{d\varphi^2} X\right\}$ where $X$ represents the
quadrupole entropy or pressure contributions of the density
fluctuations. The assembling of the second order time derivative
is done during the post-processing stage. The time derivatives are
then evaluated with a second order central differences scheme, while
the angular derivatives have been computed directly in the code to
take advantage of the spectral representation.

It is convenient to introduce the {\it characteristic strain}
\begin{equation}
h_{\rm char} (f)= \frac{1}{D} \sqrt{\frac{2}{\pi^2} \frac{G}{c^3} \frac{dE_{\rm GW}}{df}},
\end{equation}
which is related to spectrum of GW energy emitted $E_{\rm GW}/df$. The latter can be expressed in terms of the mass multipoles
\begin{equation}
\frac{d E_{\rm GW}}{df} (f) =
 \frac{G}{c^5} \frac{16\pi}{75}  (2\pi f)^2
 \sum_{m=-l}^{m=+l } \left | \widehat{\ddot Q}_{2m} \right|^2,
\end{equation}
where $\widehat{\ddot Q}_{lm}$ is the Fourier transform of $\ddot Q_{lm}$.

\section{Results}
\label{sec:results}

\subsection{Representative cases}
    \label{s:representative}

\begin{figure*}
\includegraphics[width=\columnwidth]{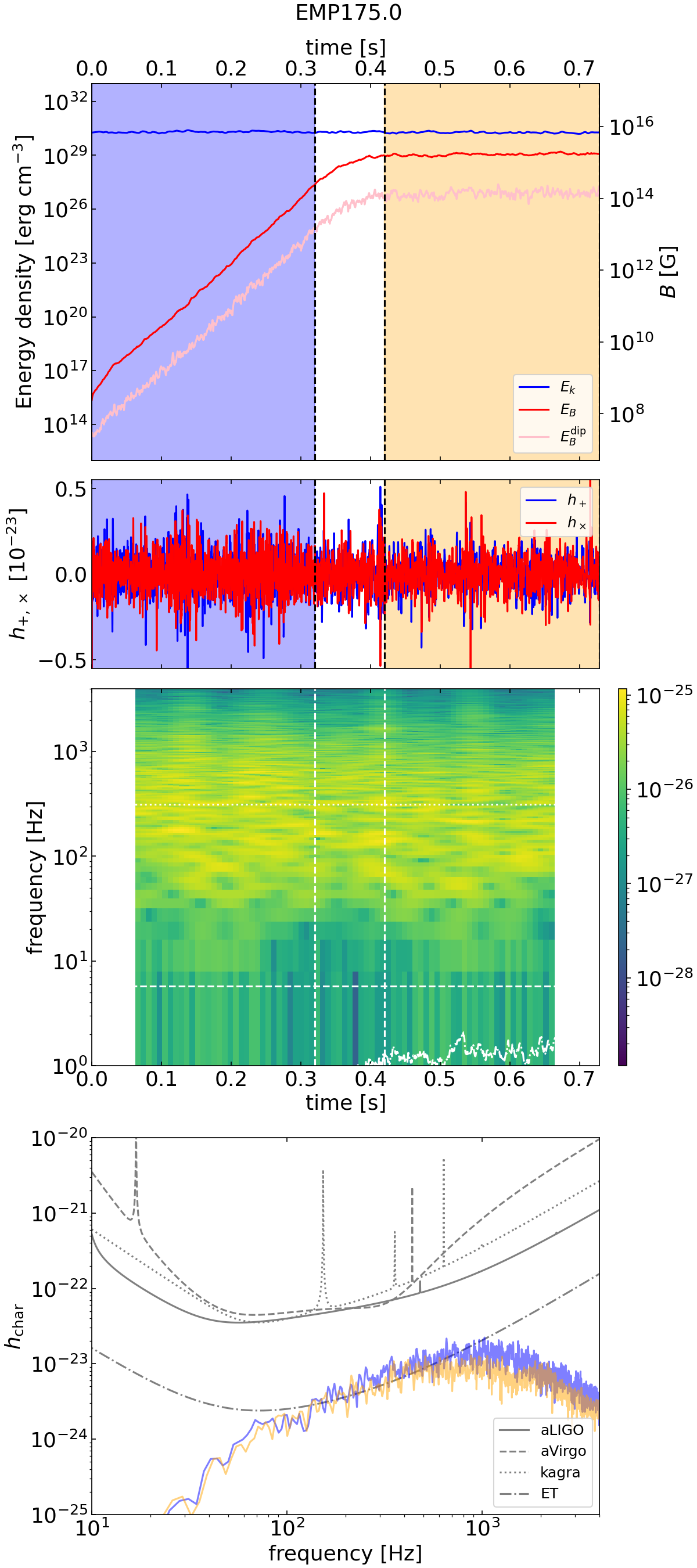}
\includegraphics[width=\columnwidth]{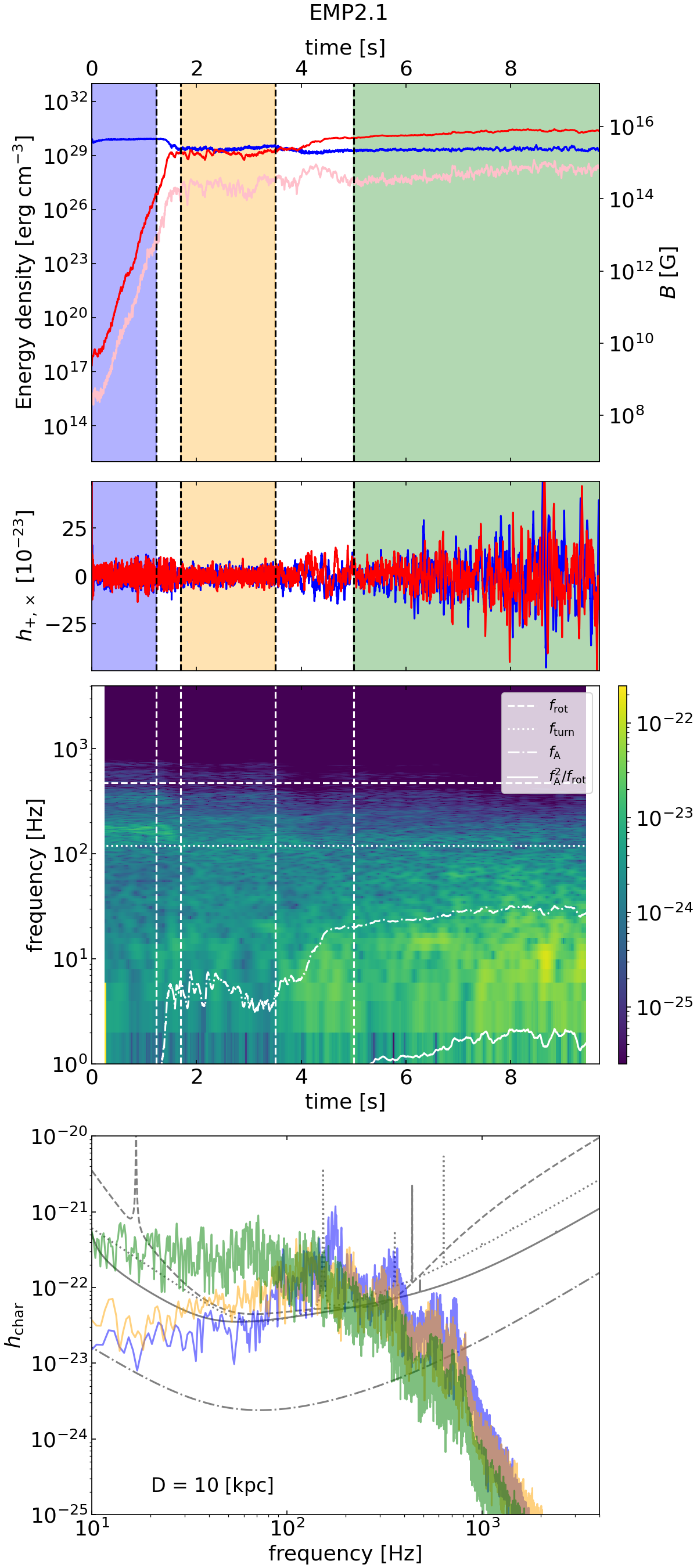}
  \caption{Summary of two typical simulations \texttt{EMP175.0} ($Pm=5$) and \texttt{EMP2.1} ($Pm=2$), respectively in the slow rotation (left panels) and the fast rotation (right panels) regime.
  The first row of panels shows the time evolution of the kinetic energy (blue line), magnetic energy (red line) and dipolar magnetic energy (pink line). The second and third rows show the GW waveform polarizations and the corresponding spectrograms (sum of both polarizations), respectively, at $10$~kpc observed along the rotation axis. Colored regions in the upper two rows correspond to the purely hydrodynamic phase (blue, defined as $E_B/E_k <10^{-3}$), the first plateau (orange) and the second plateau where the strong dynamo is active (green, only in the second model).
  Typical frequencies are overplotted on the spectrogram: the rotational frequency, $f_{\rm rot}$, the turnover frequency $f_{\rm turn}$ and the Alfv\'{e}n frequency $f_{\rm A}$.
  The bottom row shows the characteristic strain of the GW signal at $10$~kpc for each of the three phases (same colors as above) compared to the sensitivity of different GW detectors.}
\label{f:spectrogram}
\end{figure*}

Figure~\ref{f:spectrogram} compares two early time models
representative of the slow and fast rotation regimes and highlights
the impact of rotation and magnetic field on the GW signal. On the
left panel showing a model rotating at $P=\SI{175}{ms}$, we see the
kinematic growth of the dynamo (blue background on the top row) that
saturates at $t\sim\SI{0.4}{s}$ (orange background). At first sight, we
do not notice any striking difference between these two phases, when
looking at the polarization strain $h_+$ and $h_\times$ time series and the
corresponding spectrograms. The characteristic strain of the signal
shows a relatively broad spectrum around \SI{1}{kHz} which falls in
the upper bandwidth of ground-based GW detectors and just above the
design sensitivity of the Einstein Telescope (far below other
detectors) for a supernova at a distance of $\sim\SI{10}{kpc}$. A closer inspection reveals that the saturation of the
dynamo tends to slightly decrease the amplitude of the GW signal, which is
most easily seen on the bottom panel where we see the blue and orange
curves showing the characteristic strain during the kinematic (blue)
and saturated (orange) phase. This feature seems consistent with the
fact that the back reaction of the Lorentz force usually leads to a
slight decrease of the kinetic energy of the flow.

The situation is quite different on the right panel of
Fig.~\ref{f:spectrogram} that shows dynamo action in a fast rotating model at $\SI{2.1}{ms}$. As reported in \citet{Raynaud2020}, the dynamo saturation process in this fast rotation regime is characterized a first
plateau (orange background) followed by a secondary growth leading to
a stronger dynamo field (green background). When compared to the slow rotating model, key differences are already visible in the kinematic phase (blue background) where the
magnetic field has no impact on the GW signal. The characteristic strain spectrum (blue curve on the bottom row) is now above the sensitivity of all the GW detectors, and it displays several peaks between \num{e2} and \SI{e3}{Hz} (also visible in the spectrogram).  At this stage, we recall that one should not draw direct conclusions regarding the
detectability of the GW signal from Fig.~\ref{f:spectrogram} since it
derives from an early time model neglecting the structural changes of
the PNS that occurs over the integration time interval. The
sensitivity curves are nevertheless useful to highlight the increase
of the signal amplitude between the slow and fast rotation regimes.
When the magnetic field enters the force balance at $t\sim \SI{1.5}{s}$
(orange background), we first observe a decrease of the amplitude of
the GW signal qualitatively similar to what occurs in the slow rotation regime but more pronounced. The spectrum remains similar, with peak frequencies that are still visible between \num{e2} and \SI{e3}{Hz}.

Interestingly, the transition to the strong field dynamo regime (green background) is associated with a significant change of the GW spectrum, with low frequencies becoming dominant (see the spectrogram and the green curve
in the bottom panel). In parallel to the appearance of this low frequency component, the strain amplitude is significantly increased.

After presenting these typical features of the GW signals generated by
PNS turbulent convection, we discuss in more details the amplitude and
frequency scalings of the signal. To that end, we use the results of
our parameter study in which we systematically varied the PNS rotation
rate, for both early and late time\footnote{Figure~\ref{f:spectrogram} shows two early time models, but the general features discussed in Sect.~\ref{s:representative} also stand for late time models. See also Fig.~\ref{f:spectrogram309} below.} models.

\subsection{Amplitude scaling}\label{s:amp}

\begin{figure}
  \includegraphics[width=\columnwidth]{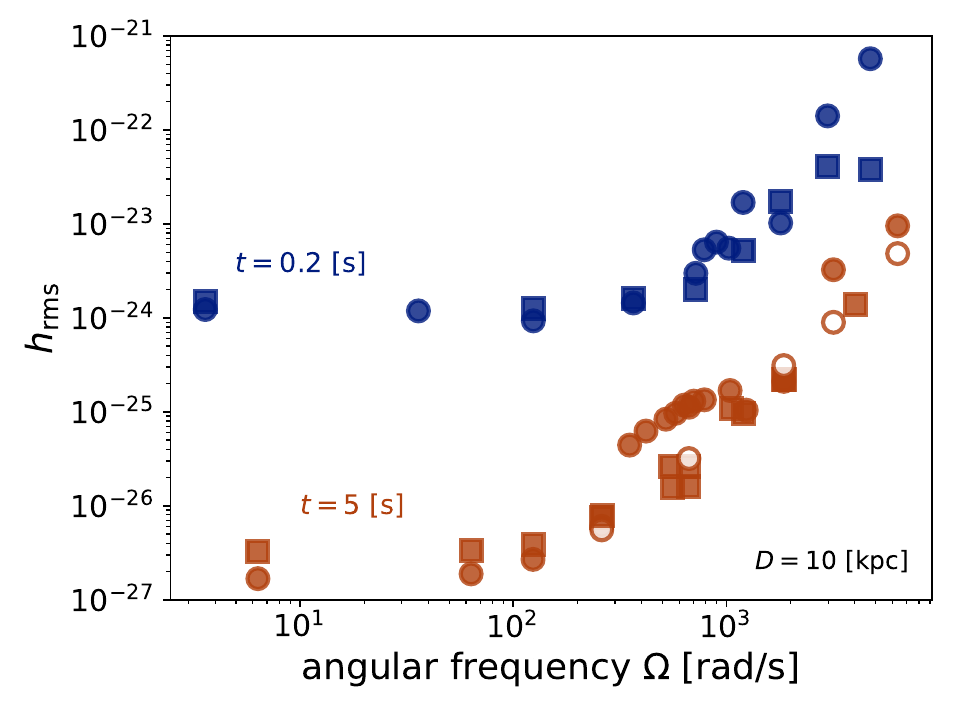}
    \caption{Gravitational wave amplitude as a function of the
      PNS rotation rate $\Omega$ for models at $t=\SI{0.2}{s}$ (blue) and
      $t=\SI{5}{s}$ (red) post-bounce.  Circles (squares) indicate
      dynamo (hydrodynamic) models. White circles indicate dynamo runs
      with a pseudo-vacuum outer boundary condition.}
    \label{f:h}
\end{figure}
\begin{figure}
    \includegraphics[width=\columnwidth]{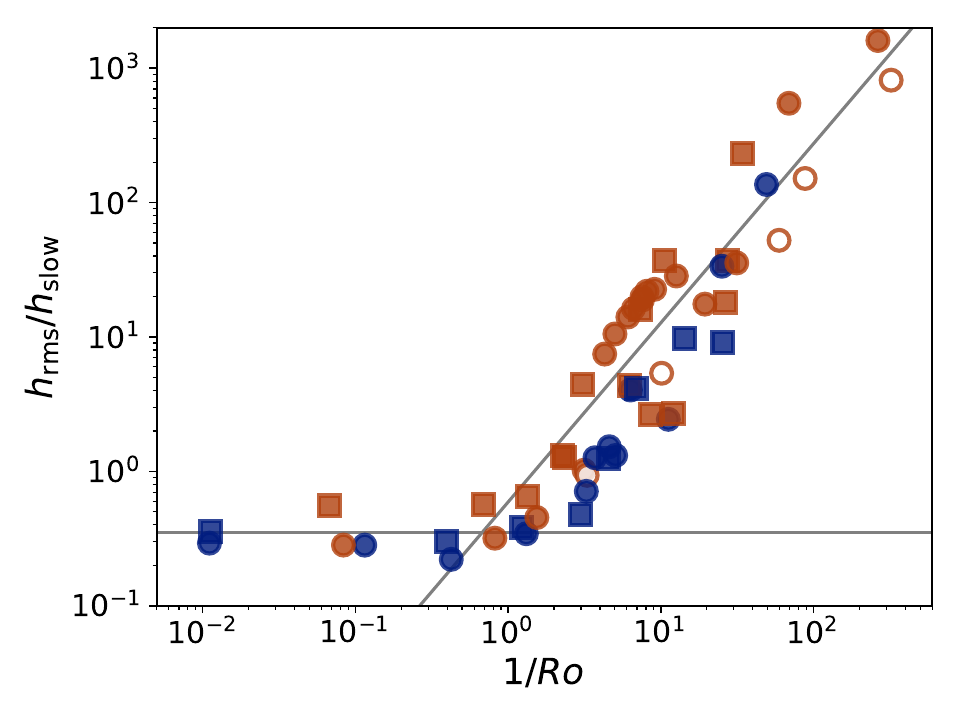}
    \caption{Normalized gravitational wave amplitude as a function of the inverse Rossby number $1/Ro$. The amplitude $h^\star=h_\text{rms}/h_\text{slow}$ has
      been rescaled using Eq.~\eqref{e:hs}. The gray lines show the
      best fits $h^\star=0.35$ (for $Ro>1$) and $h^\star= 0.59
      Ro^{-4/3}$ (for $Ro<1$). The meaning of the symbols is defined in
      Fig.~\ref{f:h}.}
    \label{f:h_norm}
\end{figure}

We define the amplitude $h_{\rm rms}$ of the GW signal as the root mean square of the strain averaged over observation angles and time (the time average being performed only on the time interval where the system converged towards its equilibrium state).
Figure~\ref{f:h} shows the evolution of $h_\text{rms}$ as a function of the PNS angular frequency for
all the models of our database.  The GW amplitude drops significantly
between early (blue) and late (red) time models. This is a direct
consequence of the decrease of the convective energy flux
$\Phi_\text{o}$ with time (see Table~\ref{tab:A}). The second
important feature is the steep increase of the rms amplitude for fast
rotation rates by up to three orders of magnitudes as compared to the
slow rotation regime. Differences between hydrodynamic (square) and
dynamo (circle) models are visible but remain minor compared to the
main trend with rotation frequency. Consistently with the description
of representative models in Sect.~\ref{s:representative},
hydrodynamic models have systematically higher GW amplitudes than the
magnetic models for $\Omega < \SI{e2}{Hz}$, and the situation reverses
for fast rotating, strong field dynamos. Note that this does not apply
to late time models with pseudo-vacuum boundary conditions (white
circles), which do not display the transition to a strong field dynamo
solution and which remain closer to the hydrodynamics cases (red
squares) and below the models with perfect-conductor boundary
conditions (red circles).  Overall, these results agree and reinforce
the conclusions drawn from the typical cases illustrated in
Fig.~\ref{f:spectrogram}. Moreover, Fig.~\ref{f:h} further suggests
the existence of two distinct regimes, since the GW amplitude becomes
independent of $\Omega$ for low rotation rates. As already observed in
\citet{Raynaud2020}, these different regimes are controlled by the
Rossby number defined by $Ro = u_\text{rms}/(\Omega d)$. This is
indeed shown in Fig.~\ref{f:h_norm}, where all the models collapse on
a master curve after rescaling the amplitudes as a function of the
inverse Rossby number $1/Ro$.

To do so, we derive the following scaling for the amplitude of the
gravitational wave based on order of magnitude considerations (and
neglecting any effects related to the magnetic field). The first step
consists in expressing the density perturbation~$\varrho'$
\begin{equation}
  \varrho' = \left.\frac{\partial
    \varrho}{\partial S}\right|_P S + \left.\frac{\partial
    \varrho}{\partial P}\right|_S P = \varrho'_S + \varrho'_P
\end{equation}
as a function of the fluid velocity.
To evaluate the entropy contribution $\varrho'_S$, we equate the
kinetic and potential energies of an entropy perturbation displaced
over a ``mixing'' length $\ell$
\begin{equation}
  \varrho'_S \tilde{g} \ell \sim \frac{1}{2} \tilde{\varrho} u^2
  \,.
  \label{e:s}
\end{equation}
To estimate the pressure contribution from an isentropic perturbation, we use the relation
\begin{equation}
  \varrho'_P = \frac{p'}{c_s^2} \sim \frac{
    \tilde{\varrho} u^2}{c_s^2}
  \,,
  \label{e:p}
\end{equation}
where we assume that the pressure perturbation is balanced by the ram
pressure. Using the hydrostatic balance ${\partial
  \tilde{P}}/{\partial r} = \tilde{\varrho} \tilde{g}$ and the
standard expression of the mixing length $\ell\sim \alpha_\text{mlt}
H_P = \alpha_\text{mlt} \left| \partial_r \ln \tilde{P}\right |^{-1}$ and introducing the adiabatic index $ \gamma = c_s^2
{\tilde{\varrho}}/{\tilde{P}}$, Eqs.~\eqref{e:s} and \eqref{e:p} give
\begin{equation}
  \frac{\varrho'}{\tilde{\varrho}}  \sim \left(\frac{\gamma}{2\alpha_\text{mlt}} +1 \right) \frac{u^2}{c_s^2}
  \,.
\end{equation}
Then, we estimate the order of magnitude of the quadrupole moments as
\begin{equation}
  Q \sim \rmid^2 M_\text{conv}\frac{\varrho'}{\tilde{\varrho}} 
  \,,
\end{equation}
where $M_\text{conv}$ is the mass of convective zone and \rmid{} the
middle radius of the convective zone. The second time derivative of
$Q$ finally gives the expression for the GW amplitude
\begin{equation}
  h \sim \frac{2G}{D c^4} \ddot{Q}
  \,,
\end{equation}
with $D$ the distance to the source.  At this point, we are left with
the estimates of the second time derivative and the fluid
velocity. For the former, we take the eddy turnover time scale $d/u $
or the inverse of the rotation rate $\Omega^{-1}$ as the typical time
scale depending on the rotation regime (see Section~\ref{s:frequency} for a description of the GW spectrum), which gives
\begin{equation}
  \ddot{Q} \sim
  \begin{cases}
    \begin{aligned}
      &Q {u^2}/{d^2}  &\text{for}\quad Ro \gg 1 \,,\\
      &Q \Omega^2 &\text{for}\quad Ro \ll 1 \,.
    \end{aligned}
  \end{cases}
\end{equation}

Finally, for the estimate of the fluid velocity, we use the
scaling laws derived by \citet{Aurnou2020} for slowly rotating and
rapidly rotating convection. They read
\begin{equation}
  u \sim   \begin{cases}
  \begin{aligned}
    &\left( \frac{\tilde{g} \tilde{\alpha} \Phi_\mathrm{o}d}{c_p 4\pi \rmid^2 \tilde{\varrho}} \right)^{1/3}
    = \left(\left|\partial_r \ln \tilde{T}\right| \frac{\Phi_\mathrm{o}}{4\pi \rmid^2 }\frac{d}{\tilde{\varrho}}  \right)^{1/3}     &\text{for}\quad Ro\gg1 \,,\\
  &\left( \left|\partial_r \ln \tilde{T}\right| \frac{ \Phi_\mathrm{o}}{4\pi \rmid^2\tilde{\varrho}} \right)^{2/5} \left(\frac{d}{2\Omega}\right)^{1/5} &\text{for}\quad Ro\ll 1
    \,.
  \end{aligned}
  \end{cases}
  \label{e:scaling_velocity}
\end{equation}
Note that a simple estimate of the convective energy flux $\Phi \sim
4\pi r^2 \tilde{\varrho} u^3$ is consistent with the slow rotating scaling
$u\propto \Phi_\mathrm{o} ^{1/3}$.

This gives the following scalings $h_\text{slow}$ and $h_\text{fast}$ for
the GW amplitude in the slow and fast rotation regimes
\begin{equation}
  h_\text{slow} \propto \frac{2G}{Dc^4} \rmid^2 M_\text{conv}
  \left(\frac{\gamma}{2\alpha_\text{mlt}} +1 \right) \frac{d^{-2/3}}{c_s^2}
   \left(\left|\partial_r \ln \tilde{T}\right|
  \frac{\Phi_\mathrm{o}}{4\pi \rmid^2 \tilde{\varrho}}
  \right)^{4/3}
  \,,
  \label{e:hs}
\end{equation}
\begin{equation}
  h_\text{fast} \propto \frac{2G}{Dc^4} \rmid^2 M_\text{conv}
  \left(\frac{\gamma}{2\alpha_\text{mlt}} +1 \right) \frac{d^{2/5}}{c_s^2}
   \left(\left|\partial_r \ln \tilde{T}\right|
  \frac{\Phi_\mathrm{o}}{4\pi \rmid^2 \tilde{\varrho}}
  \right)^{4/5}  \Omega^{8/5}
  \,.
  \label{e:hf}
\end{equation}
For the mixing length parameter, we use the typical value obtained from the calibration of field stars $\alpha_\text{mlt}\sim
2$ \citep{Valle2019}. We recall that the width of the convective
zone is related to the shell aspect ratio by the relation $d = \ro
(1-\chi)$. We estimate $c_s$, $\gamma$, $\tilde{\varrho}$, $\partial_r
\ln \tilde{T}$ at $r=\rmid$ using the background model and the EoS
LS220nl from the CompOSE
database\footnote{http://compose.obspm.fr}. Finally, $M_\text{conv}$
is directly taken from the 1D model.

Using these scalings, it is possible to rescale the numerical results at different times to a unique curve as a function of the Rossby number. The velocity scaling in the fast rotation regime (Eq.~\ref{e:scaling_velocity}) gives the following scaling for the Rossby number
\begin{equation}
  Ro = \frac{u}{d\Omega} \sim \left( \left|\partial_r \ln \tilde{T}\right| \frac{ \Phi_\mathrm{o} d}{4\pi \rmid^2\tilde{\varrho}} \right)^{2/5} \left(\frac{1}{2d^6\Omega^6}\right)^{1/5}
    \,.
  \label{e:scaling_rossby}
\end{equation}
The GW amplitude scaling can then be recast as a function of the Rossby number and the GW amplitude in the slow rotation regime
\begin{equation}
  h_\text{fast} \propto h_\text{slow} Ro^{-4/3}
  \,.
  \label{e:hf_rossby}
\end{equation}
Equation~\eqref{e:hs} is used to rescale the GW amplitudes in Fig.~\ref{f:h_norm} and we see that the transition toward the rotation dominated regime occurs around $Ro \sim1$ as expected.
The figure also shows that the scaling that we propose explains the differences in the GW amplitude between early and late time models. This confirms that the decrease of the GW amplitude is dominated by a decrease of the fluid velocity caused by a decrease of the energy flux $\Phi_\mathrm{o}$ by one order of magnitude and a similar increase of the density $\tilde{\varrho}$ (see Table~\ref{tab:A}).

\subsection{Frequency scaling}
\label{s:frequency}

\subsubsection{Slow rotation}

\begin{figure}
  \includegraphics[width=\columnwidth]{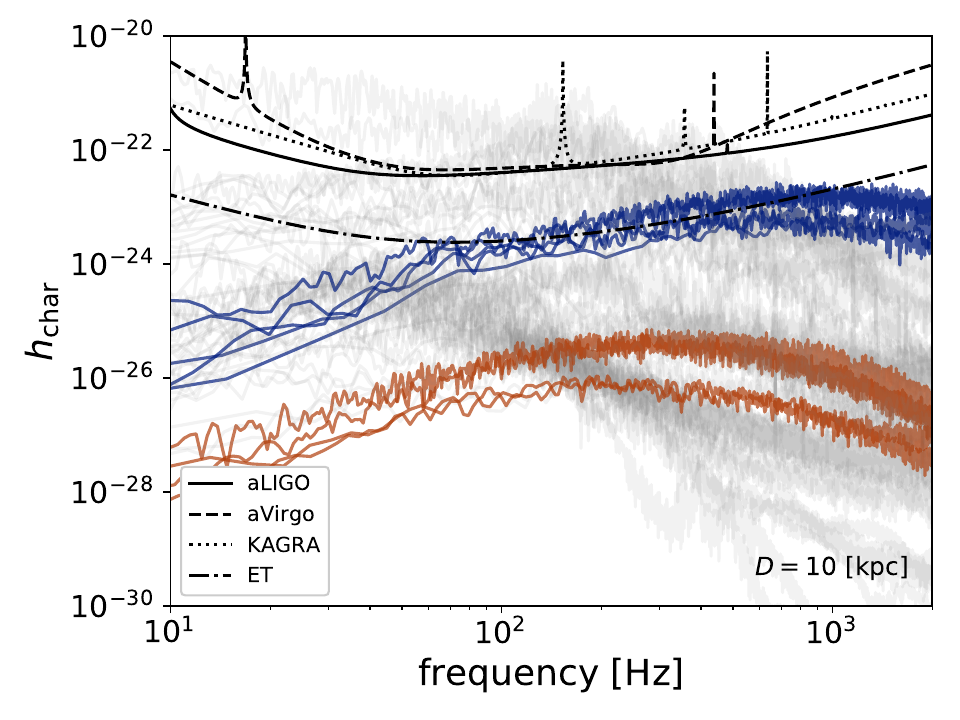}
  \includegraphics[width=\columnwidth]{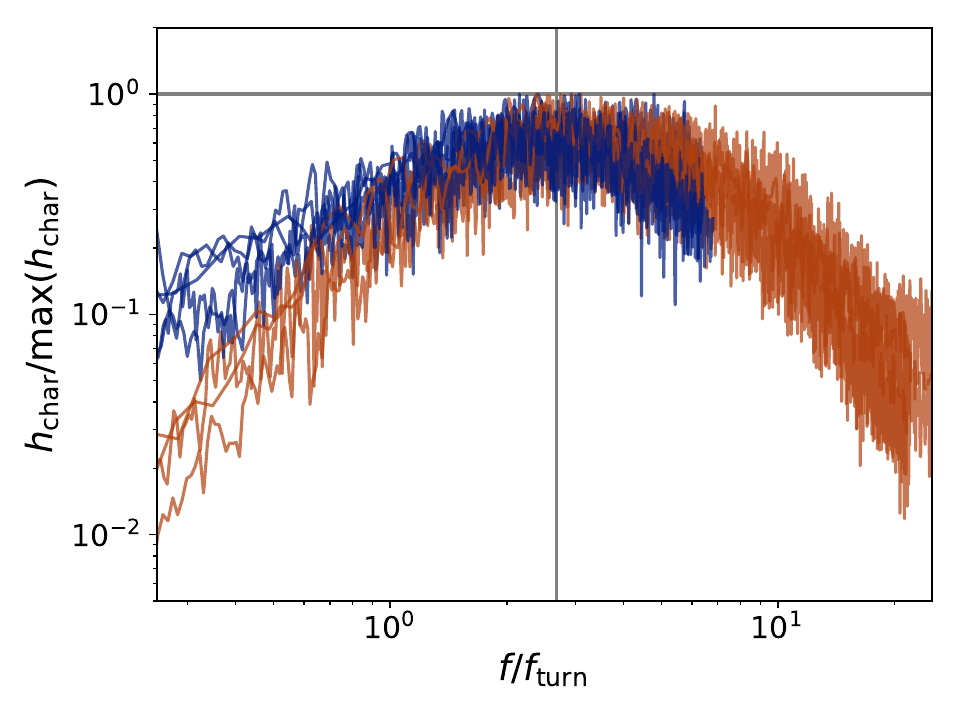}
    \caption{Top: spectrum of $h_\text{char}$ in the frequency range
      $[10,2000]$~\si{Hz}. Color lines show slow rotating models
      ($Ro>1$) at $t=\SI{0.2}{s}$ (blue) and $t=\SI{5}{s}$
      (red). Faster rotating models with $Ro<1$ are displayed in the
      grey background. The black lines show the sensitivity of
      different GW detectors. Bottom: $h_\text{char}$ scaled by its
      maximum as a function of the frequency scaled by the turnover
      frequency $f_\text{turn}$ for slow rotating models. The vertical
      line indicates the typical location of the maximum of these spectra $f/f_{\rm turn}=2.68$.}
    \label{f:slow}
\end{figure}

\begin{figure}
    \includegraphics[width=\columnwidth]{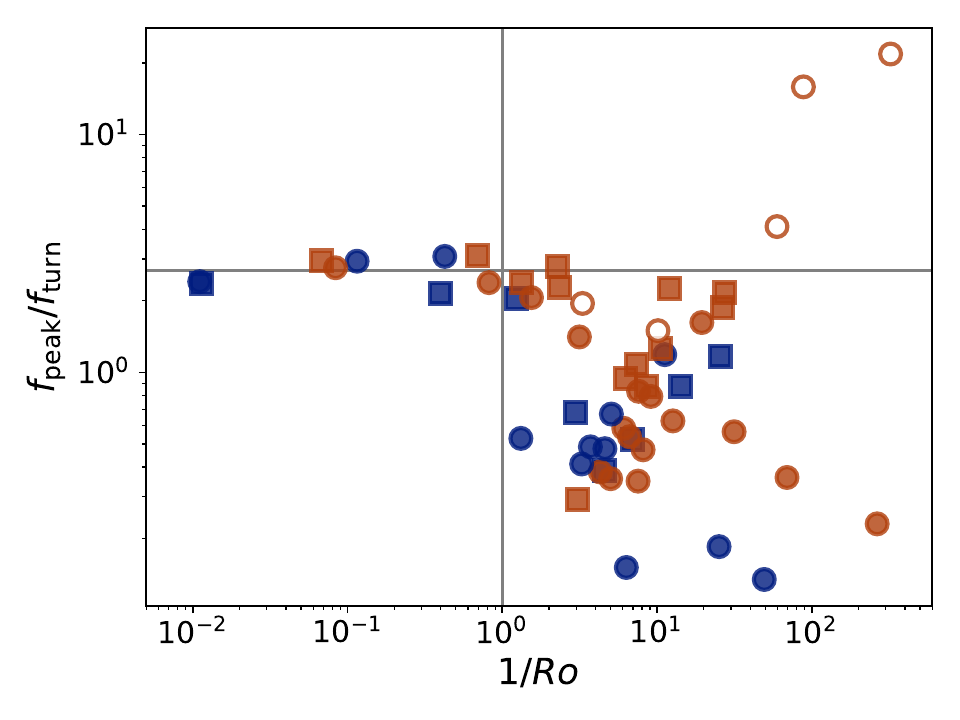}
    \caption{Peak frequency scaled by the turnover frequency
      $f_\text{turn} = u_\text{rms}/d$ as a function of $1/Ro$. The
      horizontal line $y=2.68$ fits the scaled peak frequencies for
      $Ro>1$. The meaning of the symbols is defined in
      Fig.~\ref{f:h}.}
    \label{f:turn}
\end{figure}

We display in the top panel of Fig.~\ref{f:slow} all the GW spectra of
our database and highlight with colors slow rotating models with
$Ro>1$. Both early (blue) and late (red) time models are characterized
by relatively broad spectra like the one displayed in the left panel
of Fig.~\ref{f:spectrogram}. Moreover, one can distinguish for both early and late time models a slight difference in amplitude which reflects the difference between hydrodynamic and MHD models (the latter having a lower characteristic strain). That being said, the global similarity of these spectra becomes clearer after normalizing the characteristic strain $h_{\rm char}$ by its maximum and
rescaling the frequency by the turnover frequency $f_\text{turn} =
u_\text{rms}/d$ (bottom panel of Fig.~\ref{f:slow}). This shows that
the turnover frequency is the characteristic time scale of the GW
emission induced by PNS convection in the limit of slow rotation. To
illustrate this point more quantitatively, Fig.~\ref{f:turn} shows
the dependence of the ratio of the peak frequency of these spectra
with the turnover frequency $f_\text{peak}/f_\text{turn}$ as function
of the inverse Rossby number $1/Ro$ for all models. For slow rotating
models ($1/Ro < 1$), this confirms that the maximum frequency is
indeed proportional to the turnover frequency (with a best fit at
$f=2.68 f_{\rm turn}$). On the other hand, we observe that the
turnover frequency is not the characteristic timescale once the flow
becomes rotationally constrained ($1/Ro$>1).

\subsubsection{Fast rotation}

\begin{figure}
  \includegraphics[width=\columnwidth]{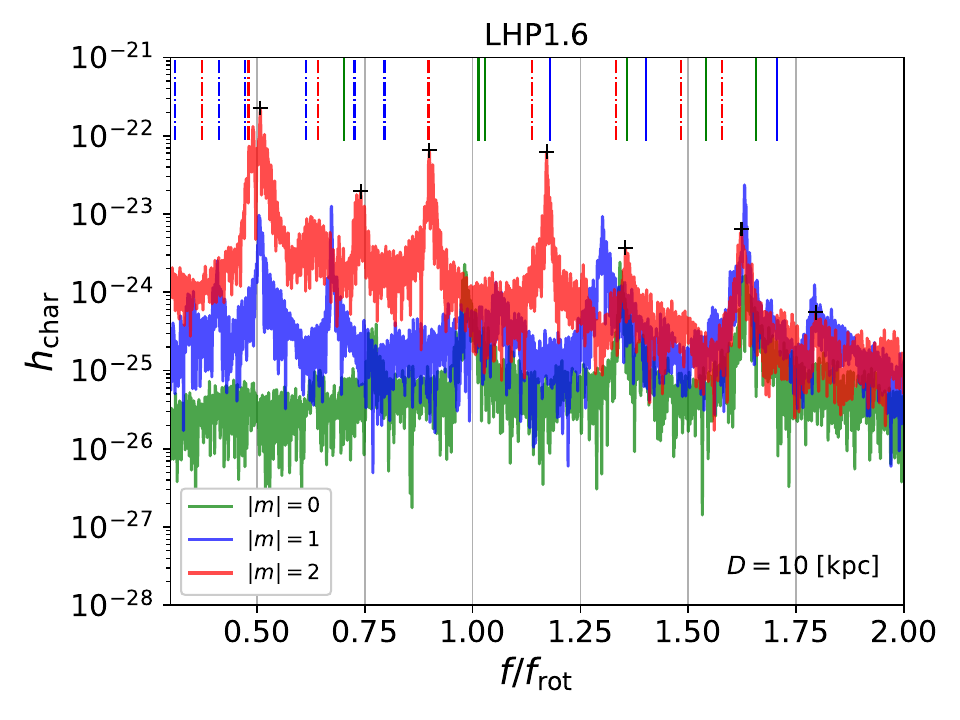}
  \includegraphics[width=\columnwidth]{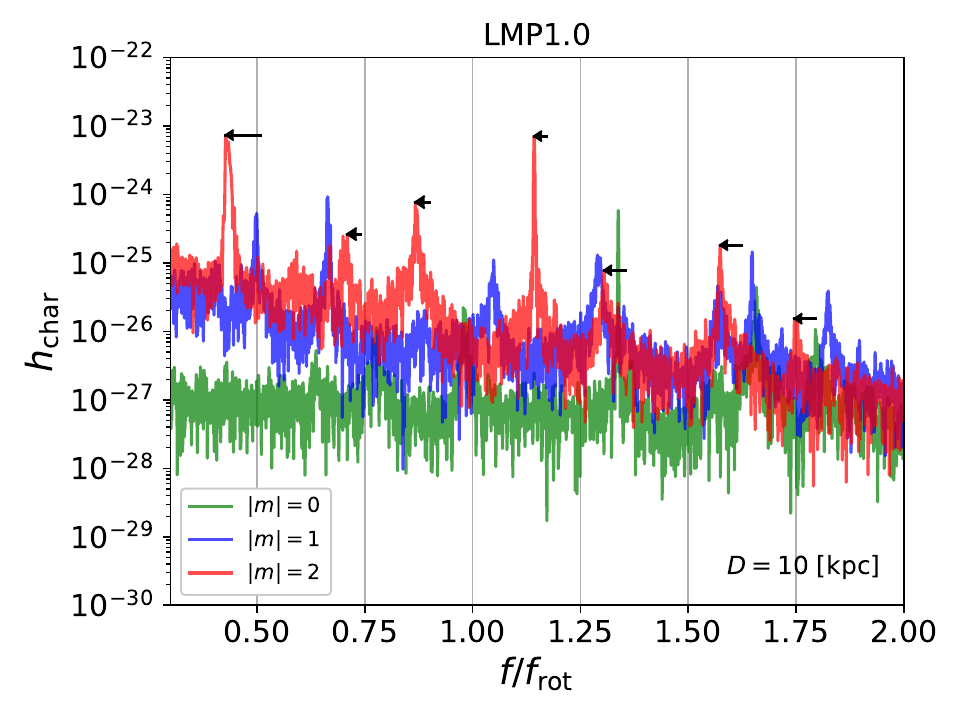}
  \caption{Characteristic strain $h_{\rm char}$ in the frequency range
    $[0.3 f_{\rm rot}- 2 f_{\rm rot}]$ for a hydrodynamic (top) and
    dynamo (bottom) model at $t=\SI{5}{s}$ post-bounce. The vertical
    lines in the top panel show inertial mode frequencies computed by
    \citet{Lockitch1999b} colored as a function of their azimuthal
    wavenumber (dashed/solid lines indicate retrograde/prograde
    modes). The black crosses in the top panel indicates some peak
    frequencies for the $|m|=2$ component and the horizontal black
    arrows highlight the frequency offset of these
    modes.}\label{f:mms}
\end{figure}

In the fast rotation regime, the gravitational wave spectra are characterized by several peaks (as already shown in the right panel of
Fig.~\ref{f:spectrogram}).  Their interpretation is complicated by the fact that these peaks originate from density fluctuations with different
azimuthal wavenumbers $|m|\in \{0,1,2\}$,
as illustrated by Fig.~\ref{f:mms}, where, for each $|m|$, we show the contributions to the waveform by the components $h_{2m}$ and $h_{2(-m)}$ (see Eq.~\eqref{eq:hlm}).
While all the three components contribute equally
to the GW emission for slow rotating models (not shown) --- which can be expected from symmetry considerations in the absence of a preferred direction ---, we find that the signal is mostly dominated by the $|m|=2$ component  for fast rotating models (red lines in Fig.~\ref{f:mms}). Other components like $|m|=1$ can also contribute significantly to the GW signal (blue lines in Fig.~\ref{f:mms}) at special frequencies where their spectrum is peaked. These peaks are clear when looking at the spectrum of a single azimuthal number but can be difficult to distinguish in the total signal combining all $m$.
A closer comparison of the top and bottom panels of Fig.~\ref{f:mms}
tends to indicate that several modes are present in both the
hydrodynamic and dynamo cases but with different relative amplitudes,
which modifies their visibility. Moreover, we note that the main peaks
of the $|m|=2$ component seem slightly shifted toward lower
frequencies (see horizontal black arrows in Fig.~\ref{f:mms}).

We claim that most of these peaks are the signature of inertial modes,
which are a class of oscillation modes that appear in rotating fluids
and for which the Coriolis force acts as the restoring force. Their
frequency in the rotating frame scales with the rotation frequency and
lies in the interval $[-2\Omega,2\Omega]$
\citep[e.g.][]{rieutord}. Hereafter we consider modes of the form $\propto e^{i(m \varphi- \omega_{\rm rot} t)}$ with positive $m$, retrogrades modes have negative frequencies while prograde modes have positives frequencies.\footnote{Note that considering only modes with positive $m$ does not exclude any physical mode because we allow negative frequencies. Indeed, a mode with negative $m$ and positive frequency $\omega_{\rm rot}$ is the same as the mode with (positive) azimuthal number $-m$ and (negative) frequency $-\omega_{\rm rot}$.}
Rossby modes (often called r-modes) are a special class of inertial
modes first studied in the dynamics of planetary atmospheres and then
in stellar radiative zones and finally neutron stars. They obey the
well-known dispersion relation
\citep[e.g.][]{Papaloizou1978,Paschalidis2017}
\begin{equation}
  \omega_{\rm rot} = \frac{2m}{l(l+1)} \Omega
  \,,\label{e:omegarot}
\end{equation}
where $\omega_{\rm rot}$ is the mode frequency in the rotating
frame. This dispersion relation applies for the case of a thin
atmosphere or zones with a strong stable stratification that prevents
radial motions. Neither condition applies to a PNS convective zone,
whose stratification is unstable and can be considered close to
isentropic (see Fig.~\ref{f:back}). \citet{Lockitch1999b} showed that
most of the Rossby modes do not exist in an isentropic star, where
only the Rossby modes with $m=l$ are present with the frequency
$\omega_{\rm rot} = 2\Omega/(m+1)$. The rest of the modes visible in the spectra are the more general class
of inertial modes; these have also been called ``rotation modes'' or
``generalised r-modes'' by \citet{Lockitch1999b}.

In order to compare inertial modes to the GW spectra computed in this
paper, their frequency should be converted to the observer frame
through the relation
\begin{equation}
  \omega_{\rm obs} = \omega_{\rm rot} - m\Omega
  \,, \label{e:rotating2observer}
\end{equation}
where $\omega_{\rm obs}$ is the mode frequency in the observer
frame. Since the modes that can contribute to GW emission have azimuthal
indices $m=0$, $1$ or $2$,
inertial modes are expected in the
range of frequencies $[0,4\Omega]$ (in absolute as measured in our spectra).
Prograde modes with $m=2$ lie in the frequency interval $[2\Omega,4\Omega]$, while retrograde modes lie in the interval $[0,2\Omega]$. For $m=1$, prograde modes lie in the interval $[\Omega,3\Omega]$, while retrograde modes lie in the interval $[0,\Omega]$.

The vertical lines in
Fig.~\ref{f:mms} show the mode frequencies obtained by
\citet{Lockitch1999b}. While this gives us an idea of the richness of
modes to be expected in this frequency range, we find that most of
them do not perfectly match with the peaks of the GW signal. However,
this discrepancy can be explained by the fact that the numerical and
the theoretical models are not built on the same hypotheses and differ
at least on these three points: first, \citet{Lockitch1999b} carried
out their computation in full sphere geometry, whereas to simplify the
numerical setup we assume a spherical shell geometry of aspect ratio
$\chi=0.1$. Second, the results we show of \citet{Lockitch1999b}
assume a polytropic equation of state of the form
$p=K\varrho^2$, which can only be considered a crude approximation of the PNS equation of state. Finally, when comparing the top and bottom panel in
Fig.~\ref{f:mms}, we see that the magnetic field could also impact the
mode frequencies, a case which has not been studied by
\citet{Lockitch1999b}.

\begin{figure}
  \includegraphics[width=\columnwidth]{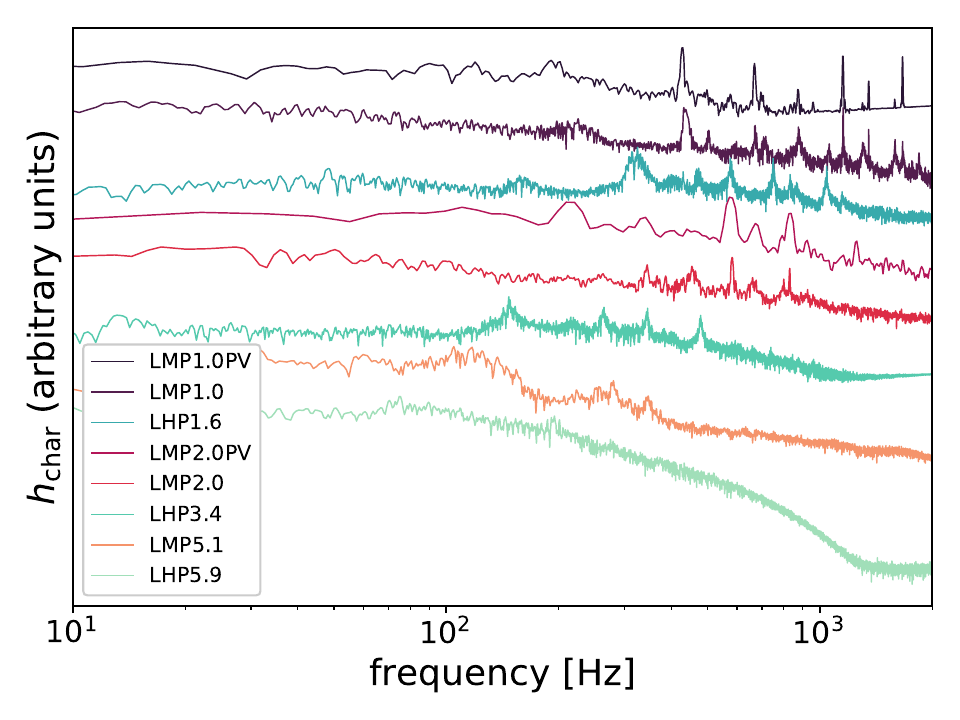}
  \includegraphics[width=\columnwidth]{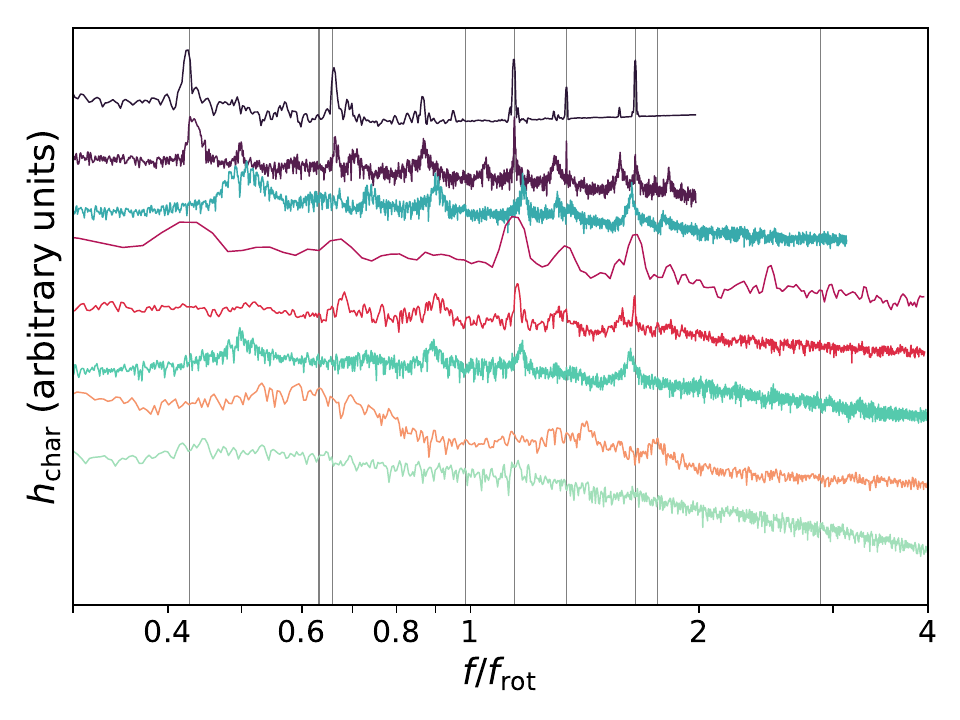}
    \caption{Top: $h_\text{char}$ (arbitrary units) in the frequency
      range $[10,2000]$~\si{Hz} for a subset of fast rotating models
      at $t=\SI{5}{s}$ post-bounce with rotation periods ranging from
      $P=\SI{1}{ms}$ to $P=\SI{5.9}{ms}$. Bottom: same quantities
      after rescaling the frequency by the rotation frequency
      $f_\text{rot}=\Omega/(2\pi)$.  The gray vertical lines show the
      frequencies of inertial modes obtained with a model of decaying
      turbulence where we fix $Ra=0$.}
    \label{f:fast}
\end{figure}

Given all these differences and the richness of modes present in the spectra, it is
difficult to identify the modes with their corresponding predictions by \cite{Lockitch1999b} probably because the uncertainty in the prediction is comparable with the separation between different modes. Our statement that these peaks are the
signature of inertial modes is rather supported by Fig.~\ref{f:fast},
where we show that the different peak frequencies scale with the
rotation rate of the PNS. In the top panel, we stack a subset of
hydrodynamic and dynamo models with rotation rates ranging from 1 to
\SI{5.9}{ms}, and whose spectra display clear peaks in the frequency
range of inertial modes. The bottom panel presents the same models
after rescaling the frequency by the rotation frequency. We also
overplot in the bottom panel the mode frequencies we obtain in a
decaying turbulence model in which we suppress the buoyancy force
setting $Ra=0$ and we initialized with an hydrodynamic turbulent state
rotating at $P=\SI{11.5}{ms}$ (see Fig.~\ref{f:ra0} in Appendix). We
see that we recover in this model most of the inertial modes obtained
in the convective cases.

Finally, we also find that the GW signal produced by fast rotating models has a circular polarization, in contrast with the slow rotating case.  A more in depth study of this aspect will be considered in an upcoming work focused on the detectability of these signals.


\subsection{A signature of the strong field dynamo}

\begin{figure}
    \includegraphics[width=\columnwidth]{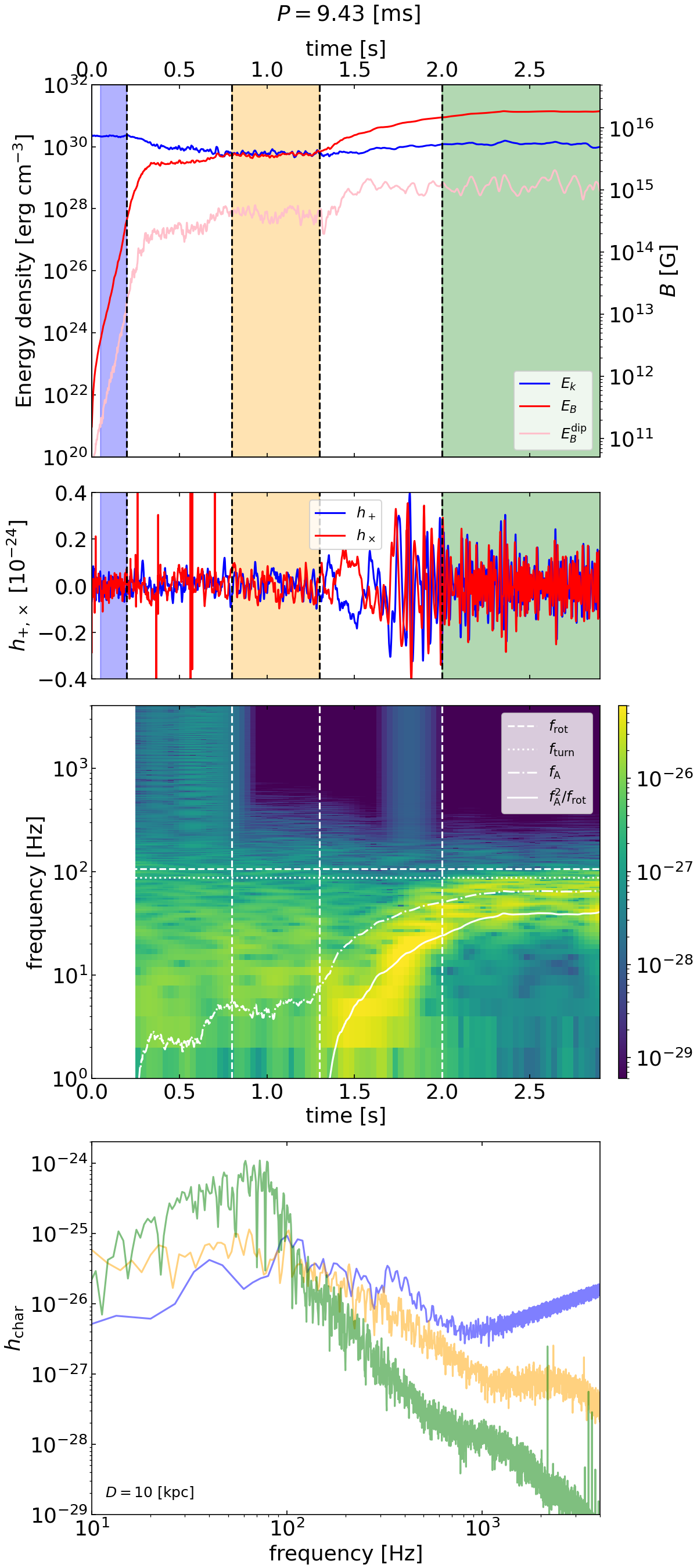}
    \caption{Summary of the late time model \texttt{LMP9.4}. The first row shows the time  evolution of the kinetic energy (blue line), magnetic energy (red line) and dipolar magnetic energy (pink line). The second and third rows show the GW waveform polarizations and the corresponding spectrogram (sum of both polarizations) at \SI{10}{kpc} observed along the rotation axis. The typical frequencies overplotted on the spectrogram are the rotational frequency $f_{\rm rot}$, the
    turnover frequency $f_{\rm turn}$, the Alfv\'{e}n frequency
    $f_{\rm A}$ and the ratio $f_{\rm A}^2/f_{\rm rot}$.
    The bottom row shows the characteristic strain of the GW signal at \SI{10}{kpc} for each of the three phases (same colors as above).}
  \label{f:spectrogram309}
\end{figure}
\begin{figure*}
    \centering
    \includegraphics[width=0.66\columnwidth]{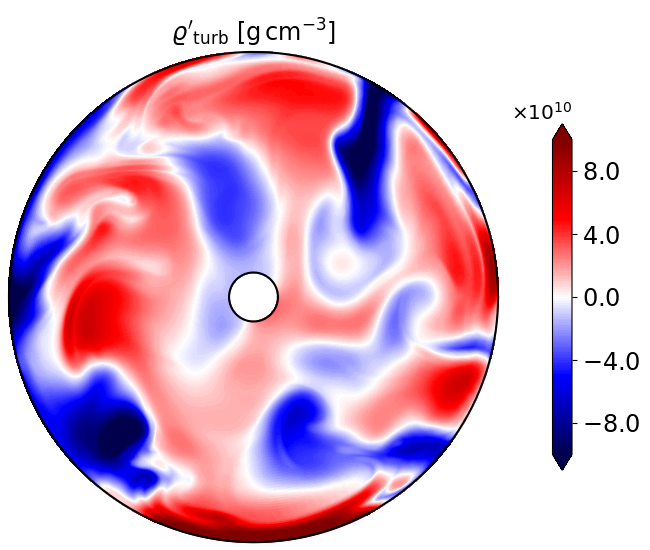}
    \includegraphics[width=0.66\columnwidth]{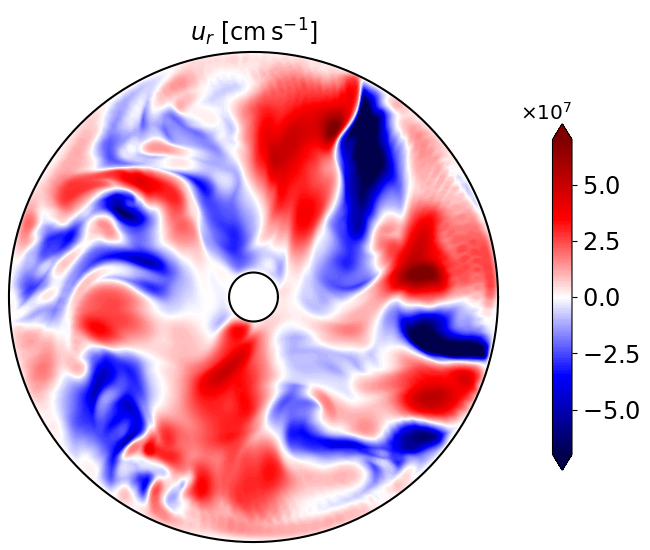}
    \includegraphics[width=0.66\columnwidth]{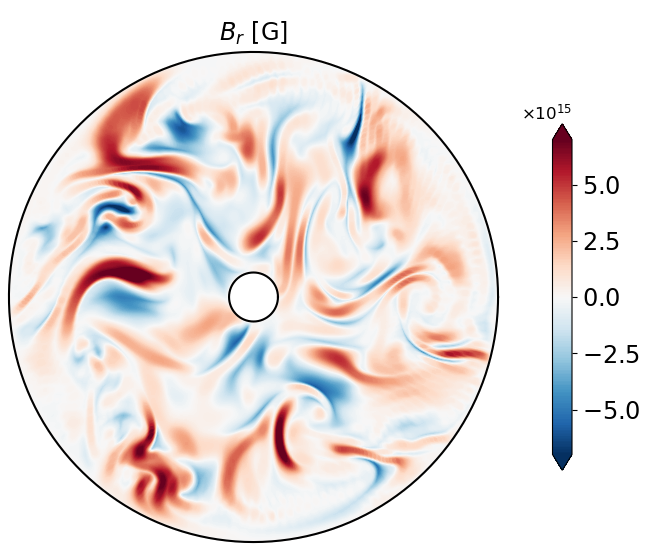}
    \includegraphics[width=0.66\columnwidth]{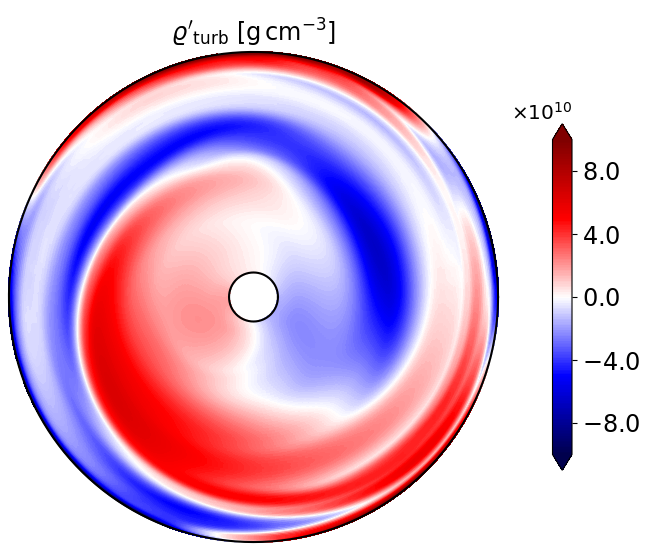}
    \includegraphics[width=0.66\columnwidth]{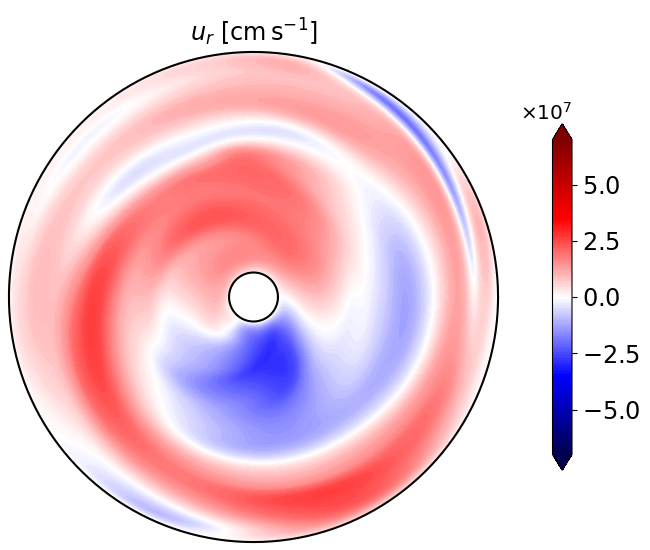}
    \includegraphics[width=0.66\columnwidth]{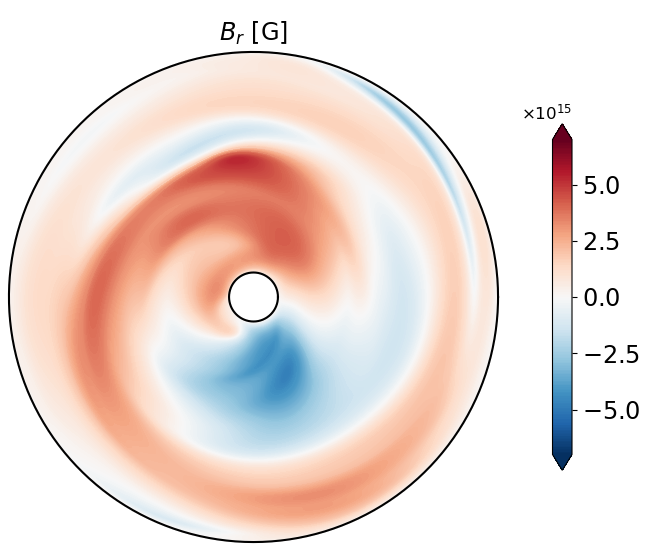}
    \caption{Equatorial slices of the non-axisymmetric density fluctuations $\varrho'_{\rm turb}$ (left), radial velocity (middle) and magnetic field (right) at $t=\SI{1.0}{s}$ (top) and $t=\SI{2.8}{s}$ (bottom) for the model \texttt{LMP9.4}.}
    \label{f:snap}
\end{figure*}
\begin{figure}
  \includegraphics[width=\columnwidth]{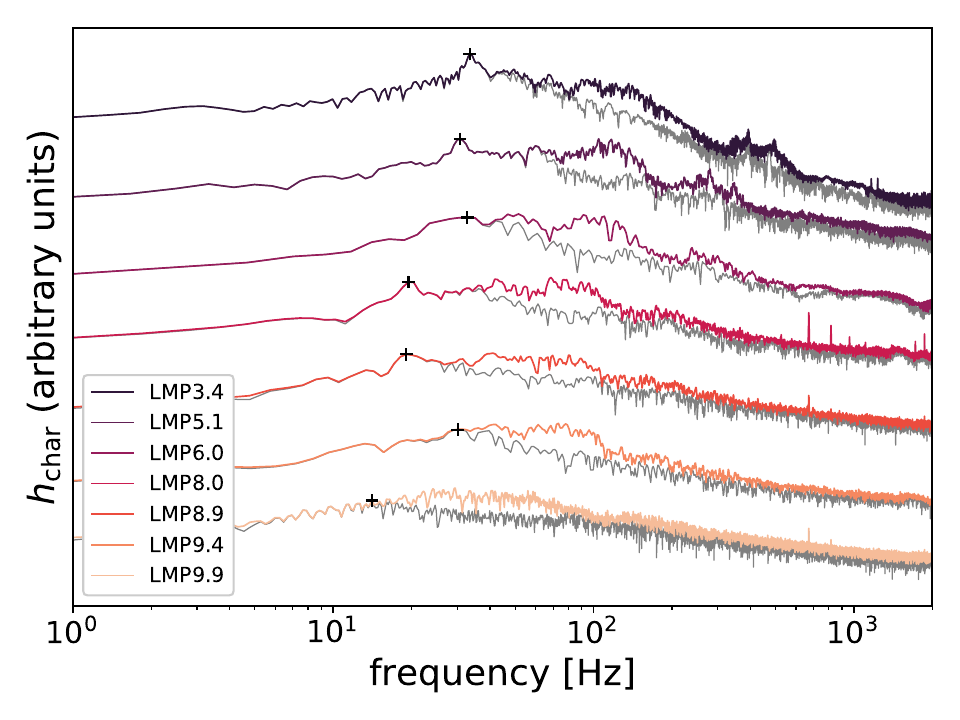}
  \caption{$h_\text{char}$ (arbitrary units) in the frequency range
    $[1,2000]$~\si{Hz} for a subset of fast rotating MHD models at
    $t=\SI{5}{s}$. The grey lines show the component of
    $h_\text{char}^{m=1}$ and the black crosses indicate its maximum.}
  \label{f:hcharB}
\end{figure}
\begin{figure}
    \includegraphics[width=\columnwidth]{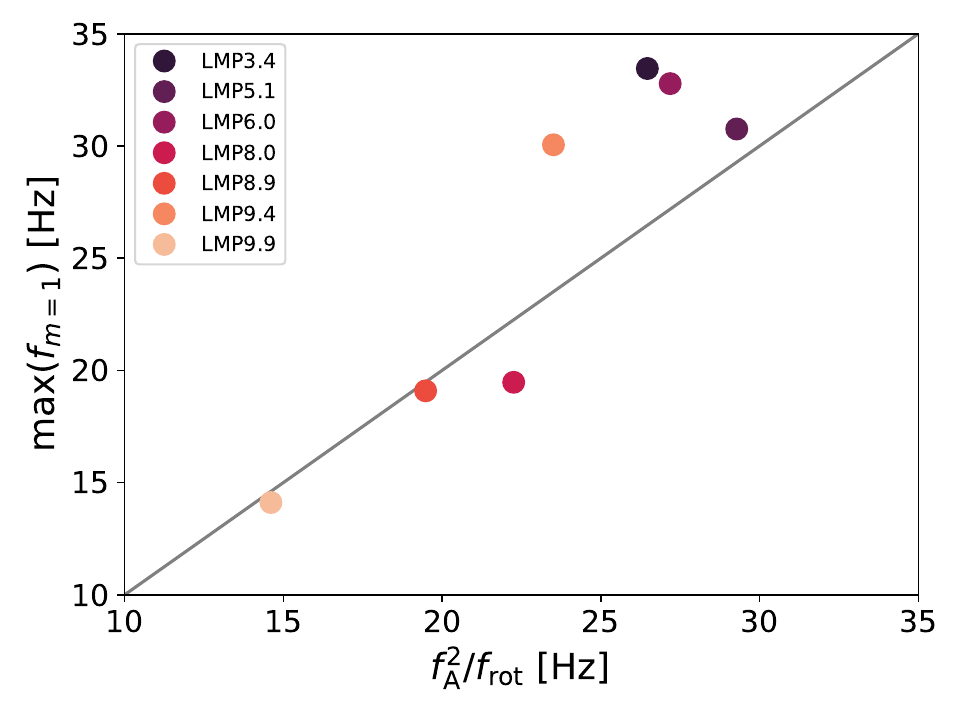}
  \caption{Peak frequency of the $m=1$ component of $h_\text{char}$ as
    a function of $f_\text{A}^2/f_\text{rot}$ for the subset of models
    displayed in Fig.~\ref{f:hcharB}}
  \label{f:scaling_mag}
\end{figure}

As discussed in Section~\ref{s:representative}, a strong GW component at low frequency appears in the fast rotating models when the
strong field dynamo sets in (see Figs.~\ref{f:spectrogram} and \ref{f:spectrogram309}). It is striking that, in a subset of our simulations, we observe a pattern with increasing frequency in time-coincidence with the increasing strength of the axisymmetric toroidal magnetic field, as we can see  in Fig.~\ref{f:spectrogram309}. Figure~\ref{f:snap} displays for the same models  snapshots of the non-axisymmetric density fluctuations and the radial component of velocity and magnetic fields, taken at the end of the first plateau ($t\sim\SI{1.0}{s}$) and in the strong field regime ($t\sim\SI{2.8}{s}$) where they become characterized by a strong $m=1$ pattern.
We therefore hypothesize that the low frequency excess is a consequence of the strong azimuthal magnetic field characteristic of the strong field dynamo branch \citep{Raynaud2020}. As shown in the previous section, the GW spectra of fast rotating modes contain a number of inertial modes. It may therefore seem natural to seek an explanation for the low frequency component in a modification of inertial modes by the azimuthal magnetic field
\citep{Hide1966,Malkus1967,Morsink2002}.

\citet{Hide1966} developed a simple analytical model of Rossby waves in a spherical shell in the presence of magnetic field. They obtained the following dispersion relation for the wave frequency in the rotating frame:
\begin{equation}
\omega_{\rm rot} \simeq \frac{\omega_{0}}{2} \pm \frac{\omega_{0}}{2}\left\lbrack 1 + 4m^2\frac{\omega_{\rm A}^2}{\omega_{0}^2} \right\rbrack^{1/2},
    \label{e:hide1}
\end{equation}
where $\omega_{0}$ is the angular frequency of the non-magnetised Rossby wave in the rotating frame, $\omega_\text{A} \equiv v_\mathrm{A}/r$ is the Alfv\'{e}n angular frequency and $m$ is the azimuthal number of the Rossby wave.  The angular frequency of the two corresponding wave solutions can be simplified in the limit $\omega_\mathrm{A}^2 \ll \omega_{0}^2 $ (which is relevant here as $\omega_\mathrm{A}^2 \lesssim 0.1 \Omega^2$) as a Rossby wave modified by magnetic effects :
\begin{equation}
\omega_{\rm rot} \simeq  \omega_{0}\left\lbrack 1 + m^2\frac{\omega_\mathrm{A}^2}{\omega_{0}^2} \right\rbrack,
\label{e:hide_rossby}
\end{equation}
and a magneto-Coriolis wave :
\begin{equation}
\omega_{\rm rot} \simeq - m^2\frac{\omega_\mathrm{A}^2}{\omega_{0}}.
\label{e:hide_MC_wave}
\end{equation}

\citet{Malkus1967} studied magneto-inertial modes in a sphere. He showed that an analytical solution can be found for a special configuration of the magnetic field: a purely toroidal magnetic field with an intensity proportional to the distance from the axis (this corresponds to a uniform current along the axis of rotation), giving rise to a uniform Alfv\'{e}n frequency in the spherical domain. Similarly to the results of \citet{Hide1966}, each inertial mode is associated to two magnetised solutions:
\begin{equation}
\omega_{\rm rot} \simeq \frac{\omega_{0}}{2} \pm \frac{\omega_{0}}{2}\left\lbrack 1 - 4m\left(\frac{\omega_0}{\Omega}+m\right)\frac{\omega_\mathrm{A}^2}{\omega_{0}^2} \right\rbrack^{1/2},
    \label{e:malkus}
\end{equation}
where $\omega_0$ is the frequency of the unmagnetised inertial mode. In the limit $\omega_A^2 \ll \omega_{0}^2 $, this simplifies once again as an inertial mode modified by magnetic effects:
\begin{equation}
\omega_{\rm rot} \simeq  \omega_{0}\left\lbrack 1 + m\left(\frac{\omega_0}{\Omega}+m\right)\frac{\omega_\mathrm{A}^2}{\omega_{0}^2} \right\rbrack,
\label{e:malkus_inertial}
\end{equation}
and a magneto-Coriolis mode:
\begin{equation}
\omega_{\rm rot} \simeq - m\left(\frac{\omega_0}{\Omega}+m\right)\frac{\omega_\mathrm{A}^2}{\omega_{0}}.
\label{e:malkus_MC_wave}
\end{equation}
Despite a difference in the numerical factor due to their different assumptions, \citet{Hide1966} and \citet{Malkus1967} agree on the conclusion that Rossby and inertial modes are modified by a magnetic field in such a way that their frequency is shifted proportionally to $ \propto {\omega_\mathrm{A}^2}/{\Omega}$ (since ${\omega_0} \propto \Omega $). Note that, in our magnetic simulations on the strong branch, the peak frequencies of $m=2$ modes tend to be systematically shifted to lower frequencies when compared to hydrodynamic simulations (see Fig.~\ref{f:mms}). This is qualitatively consistent with the expectations from \citet{Hide1966} and \citet{Malkus1967} but we do not attempt to perform a precise quantitative analysis because of the large uncertainty in measuring precisely the frequency shift from the spectra.

When interpreting the GW spectra, the mode frequency should be converted to the observer frame using Eq.~(\ref{e:rotating2observer}). With $\omega_\mathrm{A}^2 \lesssim 0.1 \Omega^2$, the magneto-Coriolis modes would appear close to the frequency $m\Omega$ with a small correction proportional to ${\omega_\mathrm{A}^2}/{\Omega}$. Since we do not observe any special features in this region of the spectrum, we conclude that magneto-Coriolis modes are not contributing significantly to the GW spectrum.

Figure~\ref{f:hcharB} shows that the low-frequency excess is predominantly generated by the $m=1$ component of the quadrupole. We therefore focus our analysis on the peak frequency of the GW spectrum generated by the $m=1$ component indicated by a cross in these spectra. For comparison with theoretical results, we evaluate the Alfv\'{e}n frequency $f_\mathrm{A}\equiv \omega_A/(2 \pi) = v_A/(2\pi r)$ by using the radius $r_{\rm mid}$ and approximating the Alfv\'{e}n speed by
\begin{equation}
  v_\mathrm{A} \simeq \frac{B_\mathrm{tor}^\mathrm{axi}}{\sqrt{4 \pi \tilde{\varrho}_\mathrm{mid}}}
  \,,
  \label{e:va}
\end{equation}
where $B_\mathrm{tor}^\mathrm{axi}$ is the rms value of the
axisymmetric toroidal magnetic field. Figure~\ref{f:scaling_mag} shows that the $m=1$ peak frequency lies very close to $f_A^2/f_{\rm rot}$. A possible interpretation is that this peak corresponds to the magnetic modification of a Rossby or inertial mode, whose frequency vanishes in the observer frame in the absence of magnetic field. Equation~(\ref{e:rotating2observer}) shows that this happens for an $m=1$ mode if $\omega_0 = \Omega$. The $l=m=1$ Rossby mode satisfies this condition and fits well in this interpretation. The observer frame frequency according to equation~(\ref{e:hide_rossby}) from \citet{Hide1966} is then $f_A^2/f_{\rm rot}$ in good agreement with Fig.~\ref{f:scaling_mag}. Note that equation~(\ref{e:malkus_inertial}) from \citet{Malkus1967} predicts the same scaling with Alfv\'{e}n and rotation frequency but a factor 2 higher. The time evolution of the low frequency mode appearing
in Fig.~\ref{f:spectrogram} (right side) and in Fig.~\ref{f:spectrogram309} also seem to follow the time evolution of $f_A^2/f_{\rm rot}$.

\section{Conclusion}
\label{sec:conclusions}

In this paper, we have investigated the gravitational waves emitted by convective motions and the associated dynamo inside a proto-neutron star. While this signal is probably subdominant before the launch of the supernova explosion, it may become the dominant source of GW in the few seconds following the explosion. We have computed GW waveforms by applying the quadrupole formula to outputs of 3D MHD numerical simulations describing the convective zone in the anelastic approximation. The simulations follow from \citet{Raynaud2020} with a PNS structure representative of early times (0.2s after bounce) as well as an extension of their setup with a PNS structure representative of late post-explosion times (5s after bounce). With both PNS structures, we performed and analysed a series of simulations in which the rotation rate is systematically varied. Our main results can be summarized as follows:
\begin{itemize}
    \item In the slow rotation regime (Rossby numbers larger than 1), convection emits gravitational waves with a broad spectrum peaking at about three times the turnover frequency. Compared to non-magnetised simulations, the magnetic field impacts the dynamics in a such a way as to decrease slightly the GW amplitude without changing significantly the spectrum.
    \item In the fast rotation regime (Rossby numbers smaller than 1), the gravitational wave amplitude increases steeply with the rotation rate. The rms strain reaches values up to a few thousands times larger than in a corresponding non-rotating simulation. The signal is also circularly polarised, unlike in the slow rotation regime.
    \item The GW spectrum of fast rotating convection is characterized by the presence of several peaks whose frequency scales with the rotation rate, which we interpret as inertial modes.
    \item The growth of the axisymmetric toroidal magnetic field characteristic of the strong field dynamo solution described by \citet{Raynaud2020} is accompanied by an increase of the GW amplitude and the appearance of a low-frequency excess in the GW spectrum at frequencies $\lesssim 100 \, \si{Hz}$. This low-frequency excess is mostly emitted by the $m=1$ component of the quadrupole, while the $m=2$ component is the main contributor to the rest of the spectrum (except for some peaks associated to inertial modes). We observe that the emission by the $m=1$ component peaks at a frequency proportional to the square of the Alfv\'{e}n frequency divided by the rotation frequency. This low-frequency emission is a signature of the strong field dynamo and is tentatively interpreted as due to the $m=1$ Rossby mode modified by the toroidal magnetic field.
    \item By using simple physical arguments, we derive scaling laws for the GW amplitude in the regime of slow and fast rotation. These scaling laws compare favorably with the numerical results.
\end{itemize}

The identification in the GW spectrum of fast rotating convection of peaks associated to inertial modes opens interesting new perspectives for PNS asterosismology. g, f and p modes have been identified previously in the GW spectrum \citep{Torres-Forne2018,Morozova2018,Torres-Forne2019b,Torres-Forne2019a,Sotani2019,Sotani2020} and have the potential to constrain PNS properties such as combinations of radius and mass \citep{Bizouard2021}. The identification of inertial modes would open the possibility to constrain the rotation frequency of the PNS. This may be a challenging task because of the richness of inertial modes potentially present in the spectra and deserves further investigation. The GW excess at low frequency appearing in the strong field dynamo opens furthermore the perspective to constrain the dynamo efficiency and the strength of the magnetic field inside the PNS.

We stress that the computation of our waveforms has several limitations: i) with our current numerical setup, we cannot compute a single continuous waveform including the evolution of the different features as the PNS contracts and cools down. Instead we have probed the features of this GW signal at different times during this cooling phase. This is sufficient to make an estimation of the typical values of the strain and frequencies but does not allow us to produce realistic GW templates. ii) The second important limitation of these simulations is that they are restricted to the convective zone of the PNS. While this is useful to isolate and study the physics of a specific signal, it misses several other important sources of GW including $g$-modes from the PNS surface and possibly the SASI. Before the onset of explosion, these other signals are probably dominant and would hide the signal from convection computed here. The relevance of our analysis may therefore be limited to times after the onset of the explosion. The time of the onset of the explosion will affect greatly the amplitude of the GW strain at the time it is first observed in a clean environment (the later the weaker). And iii) this work is limited to a particular PNS structure obtained from a 1D simulation based on a specific progenitor and with the LS220 EOS to define the anelastic background. \cite{Roberts2012} has shown that the duration of the PNS convection is very sensitive to the nuclear symmetry energy, so different EOS could produce a longer or shorter duration signal. It also remains to be explored how the amplitude of the GW signal depends on the EOS (and other microphysical properties such as the neutrino opacities) and on the progenitor structure.

Although the contribution from PNS convection to the GW strain of CCSNe is in general subdominant (with respect to the pre-explosion contribution), the typical strain and frequency range and the potentially long duration of the signal, could allow its detection by current GW detectors in a nearby supernova explosion, and may be a primary target for next generation of GW detectors, the Einstein Telescope \citep{Hild2011}  and Cosmic Explorer \citep{Reitze2019}. The detectability of these signals is out of the scope of this work and will be considered in an upcoming work.

Further work should also extend the numerical domain to include other regions of the PNS outside of the convective zone. This would in particular allow us to study the interaction of the convective dynamo with the magnetorotational instability which develops in the stably stratified outer layers of the PNS \citep[e.g.][]{Guilet2015b,Reboul-Salze2021} and with the $g$-mode oscillations of the PNS. It will also be interesting to study how this signal could compare to other important GW signal contributors after the onset of explosion, like the aspherical explosion morphology \citep{Obergaulinger2006,Scheidegger2010} and anisotropic neutrino emission \citep{Muller2012,Vartanyan2020}, with typical frequencies of $1-\SI{10}{Hz}$, that may lay in a similar frequency range as the $m=1$ signature of the dynamo observed in our simulations.

\section*{Acknowledgements}
RR and JG acknowledge support from the European Research Council
(MagBURST grant 715368). PCD acknowledges the support from the grants PGC2018-095984-B-I00, PROMETEU/2019/071 and the {\it Ramon y Cajal} funding (RYC-2015-19074) supporting his research. We thank T. Gastine for his help with numerical developments and thermodynamic relations. We thank T. Janka for sharing the 1D model data. We thank M. Bugli, T. Foglizzo, B. Gallet, V. Prat and
A.~Srugarek for discussions and comments and J. de Cabo for checking some of the complicated equations. The idea of this project emerged during the
program ``Gamma-ray bursts and supernovae: from the central engines to
the observer'' of the PSI2 project funded by the IDEX Paris-Saclay,
ANR-11-IDEX-0003-02.  Numerical simulations have been carried out at
the CINES on the Occigen supercomputer (DARI projects A0070410317 and
A0090410317).

\section*{Data Availability}
The code MagIC is freely available online at \url{https://magic-sph.github.io}. The simulation outputs are available upon request.


\bibliographystyle{mnras}
\bibliography{export-bibtex,raynaud}



\appendix

\section{Numerical models}

\begin{figure}
    \includegraphics[width=\columnwidth]{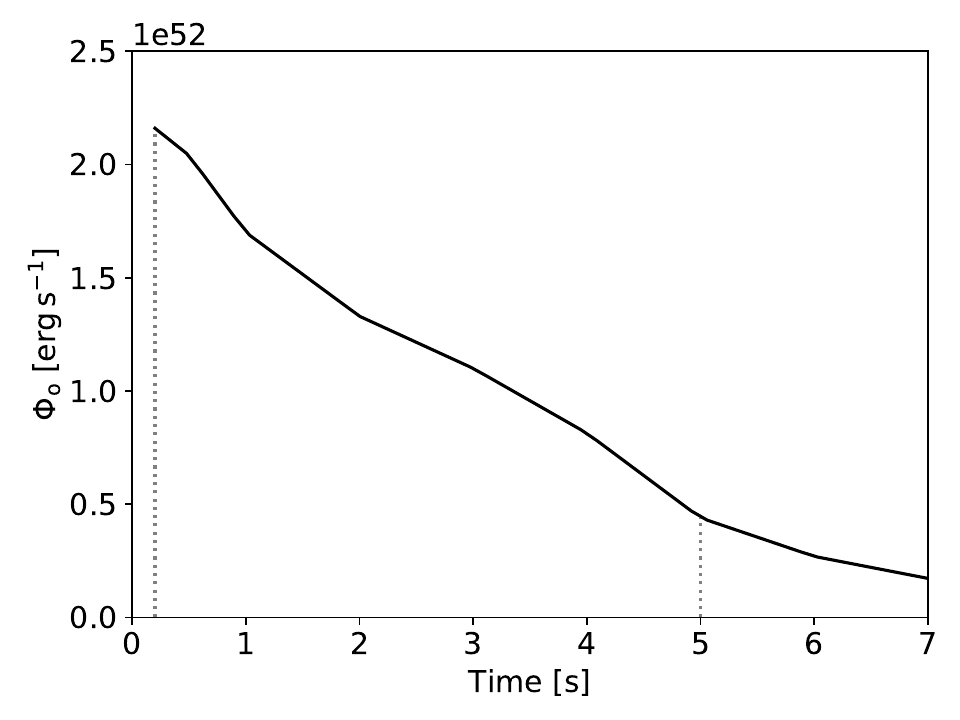}
    \caption{Post-bounce evolution of $\Phi_{\rm o}$ as a function of time according to our 1D background model. The vertical dotted line indicate the time $t=\SI{0.2}{s}$ and $t=\SI{5}{s}$ corresponding to the early and late time models.}
    \label{f:phi_o}
\end{figure}

\begin{figure}
  \includegraphics[width=\columnwidth]{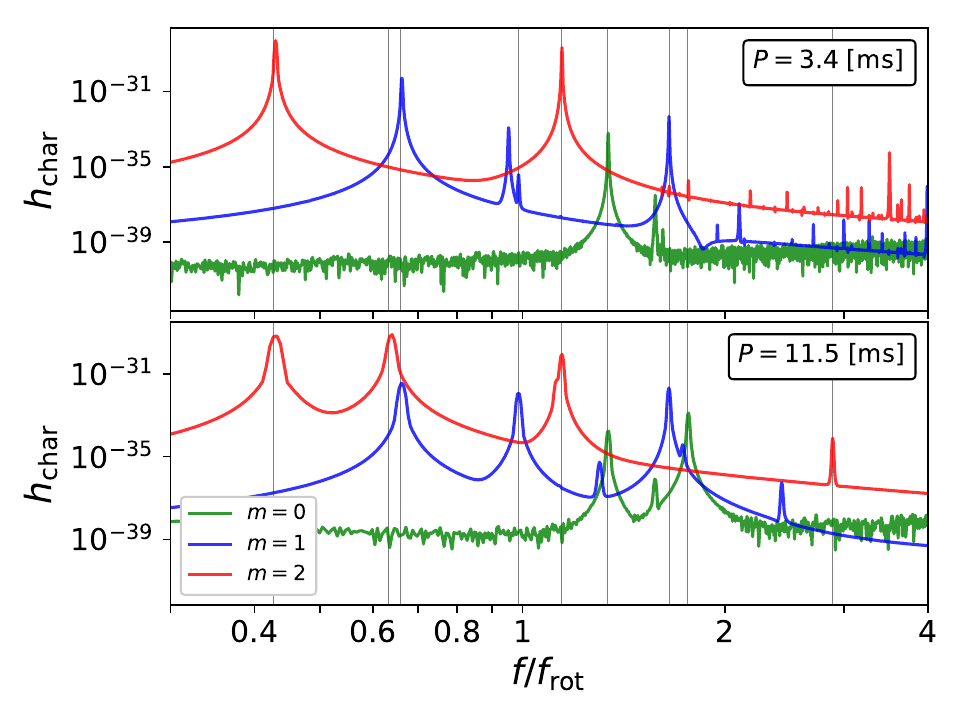}
  \caption{Spectra of models initialized with late time PNS models in
    a turbulent state and evolved in time without the buoyancy
    force. The periods indicate the rotation period of turbulent model
    used for the initial condition. The modes highlighted with
    vertical grey lines are the ones reported in
    Fig.~\ref{f:fast}.}\label{f:ra0}
\end{figure}

Figure~\ref{f:phi_o} shows the time evolution of the energy flux $\Phi_{\rm o}$. 
Figure~\ref{f:ra0} shows the characteristic strains of
two models initialized with hydrodynamic turbulent states in which we suppress the buoyancy force setting $Ra=0$. 
Finally, Table~\ref{tab:models} summarizes all the numerical simulations used in this work.


\begin{table*}
  \centering
  \caption{Overview of the numerical simulations carried out. The
    thermal Prandtl number is fixed to $Pr=0.1$ for all models. The
    names follow this convention: the first letter differentiates
    early (E) and late (L) time models, the second one hydrodynamic
    (H) or magnetic (M) models and then rotation period. The majority
    of models use a perfect conductor outer boundary condition, except
    the ones whose name end 'PV' refers to pseudo-vacuum boundary
    condition. Early time models ($t=\SI{0.1}{s}$ post-bounce) have an
    aspect ratio~$\chi=0.5$ whereas late time models ($t=\SI{5}{s}$
    post-bounce) have an aspect ratio~$\chi=0.1$. The MagIC code uses
    a slightly different definition of the Rayleigh number and we give
    here the numerical values used in code namelist. $B_0$ corresponds to the initial magnetic field strength. $\Delta T$ indicates the duration of the time interval over which time averaging is performed once the system reaches a steady state. $N_r$ and
    $N_\phi$ correspond to the number of radial and azimuthal grid
    points, respectively. The corresponding spectral resolution for
    the spherical harmonic decomposition is given by
    $\ell_\text{max}=\lfloor N_{\phi}/3 \rfloor$.}
  \label{tab:models}
  \tabcolsep=0.1cm
  \small
  \begin{tabular}{lccccccccccccr}
    \toprule
     Name  &  $Pm$  &  $Ra^\mathrm{M}$ &  $E$  &  $P$ [ms]  &  $Ro$  &  $h_\mathrm{rms}$  &  $h_\mathrm{slow}$  &  $f_\mathrm{peak}$ [Hz]  &  $f_\mathrm{turn}$ [Hz]  &  $B_0$ [G]  &  $\Delta T$ [s]  &  $N_r$  &  $N_\phi$ \\
    \midrule
\texttt{EHP1.3} &  & $ 1.77 \times 10^{4} $ & $ 5.00 \times 10^{-4} $ & $ 1.33 \times 10^{0} $ & $ 3.92 \times 10^{-2} $ & $ 3.82 \times 10^{-23} $ & $ 3.90 \times 10^{-24} $ & $ 2.17 \times 10^{2} $ & $ 1.85 \times 10^{2} $ &   & $ 0.42 $ & $145$ & $320$ \\
\texttt{EMP1.3} & $2$ & $ 1.77 \times 10^{4} $ & $ 5.00 \times 10^{-4} $ & $ 1.33 \times 10^{0} $ & $ 2.07 \times 10^{-2} $ & $ 5.72 \times 10^{-22} $ & $ 3.90 \times 10^{-24} $ & $ 1.32 \times 10^{1} $ & $ 9.80 \times 10^{1} $ & $ 9.2 \times 10^{15} $ & $ 4.19 $ & $145$ & $320$ \\
\texttt{EHP2.1} &  & $ 8.84 \times 10^{3} $ & $ 1.00 \times 10^{-3} $ & $ 2.11 \times 10^{0} $ & $ 7.01 \times 10^{-2} $ & $ 4.08 \times 10^{-23} $ & $ 3.90 \times 10^{-24} $ & $ 1.83 \times 10^{2} $ & $ 2.09 \times 10^{2} $ &   & $ 1.01 $ & $145$ & $320$ \\
\texttt{EMP2.1} & $2$ & $ 8.84 \times 10^{3} $ & $ 1.00 \times 10^{-3} $ & $ 2.11 \times 10^{0} $ & $ 4.01 \times 10^{-2} $ & $ 1.41 \times 10^{-22} $ & $ 3.90 \times 10^{-24} $ & $ 2.21 \times 10^{1} $ & $ 1.19 \times 10^{2} $ & $ 2.4 \times 10^{9} $ & $ 2.32 $ & $145$ & $320$ \\
\texttt{EHP3.5} &  & $ 8.84 \times 10^{3} $ & $ 1.66 \times 10^{-3} $ & $ 3.50 \times 10^{0} $ & $ 1.45 \times 10^{-1} $ & $ 1.75 \times 10^{-23} $ & $ 3.90 \times 10^{-24} $ & $ 1.36 \times 10^{2} $ & $ 2.60 \times 10^{2} $ &   & $ 4.65 $ & $181$ & $512$ \\
\texttt{EMP3.5} & $5$ & $ 8.84 \times 10^{3} $ & $ 1.66 \times 10^{-3} $ & $ 3.50 \times 10^{0} $ & $ 8.95 \times 10^{-2} $ & $ 1.02 \times 10^{-23} $ & $ 3.90 \times 10^{-24} $ & $ 1.91 \times 10^{2} $ & $ 1.61 \times 10^{2} $ & $ 2.5 \times 10^{9} $ & $ 1.51 $ & $181$ & $512$ \\
\texttt{EHP5.3} &  & $ 8.84 \times 10^{3} $ & $ 2.49 \times 10^{-3} $ & $ 5.25 \times 10^{0} $ & $ 2.19 \times 10^{-1} $ & $ 5.24 \times 10^{-24} $ & $ 3.90 \times 10^{-24} $ & $ 1.01 \times 10^{2} $ & $ 2.61 \times 10^{2} $ &   & $ 0.77 $ & $201$ & $864$ \\
\texttt{EMP5.3} & $5$ & $ 8.84 \times 10^{3} $ & $ 2.49 \times 10^{-3} $ & $ 5.25 \times 10^{0} $ & $ 1.59 \times 10^{-1} $ & $ 1.69 \times 10^{-23} $ & $ 3.90 \times 10^{-24} $ & $ 2.88 \times 10^{1} $ & $ 1.90 \times 10^{2} $ & $ 5.8 \times 10^{15} $ & $ 0.80 $ & $201$ & $864$ \\
\texttt{EMP6.1} & $5$ & $ 8.84 \times 10^{3} $ & $ 2.91 \times 10^{-3} $ & $ 6.13 \times 10^{0} $ & $ 1.98 \times 10^{-1} $ & $ 5.51 \times 10^{-24} $ & $ 3.90 \times 10^{-24} $ & $ 1.36 \times 10^{2} $ & $ 2.03 \times 10^{2} $ & $ 2.6 \times 10^{15} $ & $ 0.60 $ & $201$ & $864$ \\
\texttt{EMP7.0} & $5$ & $ 8.84 \times 10^{3} $ & $ 3.32 \times 10^{-3} $ & $ 7.00 \times 10^{0} $ & $ 2.18 \times 10^{-1} $ & $ 6.40 \times 10^{-24} $ & $ 3.90 \times 10^{-24} $ & $ 9.37 \times 10^{1} $ & $ 1.96 \times 10^{2} $ & $ 5.6 \times 10^{15} $ & $ 0.24 $ & $241$ & $864$ \\
\texttt{EMP8.0} & $5$ & $ 8.84 \times 10^{3} $ & $ 3.79 \times 10^{-3} $ & $ 7.99 \times 10^{0} $ & $ 2.70 \times 10^{-1} $ & $ 5.30 \times 10^{-24} $ & $ 3.90 \times 10^{-24} $ & $ 1.03 \times 10^{2} $ & $ 2.13 \times 10^{2} $ & $ 5.2 \times 10^{15} $ & $ 0.39 $ & $241$ & $864$ \\
\texttt{EHP8.8} &  & $ 8.84 \times 10^{3} $ & $ 4.15 \times 10^{-3} $ & $ 8.75 \times 10^{0} $ & $ 3.36 \times 10^{-1} $ & $ 2.01 \times 10^{-24} $ & $ 3.90 \times 10^{-24} $ & $ 1.64 \times 10^{2} $ & $ 2.41 \times 10^{2} $ &   & $ 2.60 $ & $181$ & $512$ \\
\texttt{EMP8.8} & $5$ & $ 8.84 \times 10^{3} $ & $ 4.15 \times 10^{-3} $ & $ 8.75 \times 10^{0} $ & $ 3.08 \times 10^{-1} $ & $ 2.98 \times 10^{-24} $ & $ 3.90 \times 10^{-24} $ & $ 9.11 \times 10^{1} $ & $ 2.21 \times 10^{2} $ & $ 4.0 \times 10^{15} $ & $ 0.23 $ & $201$ & $512$ \\
\texttt{EHP17.2} &  & $ 8.84 \times 10^{3} $ & $ 8.15 \times 10^{-3} $ & $ 1.72 \times 10^{1} $ & $ 8.11 \times 10^{-1} $ & $ 1.60 \times 10^{-24} $ & $ 3.90 \times 10^{-24} $ & $ 6.04 \times 10^{2} $ & $ 2.97 \times 10^{2} $ &   & $ 0.48 $ & $201$ & $864$ \\
\texttt{EMP17.2} & $5$ & $ 8.84 \times 10^{3} $ & $ 8.15 \times 10^{-3} $ & $ 1.72 \times 10^{1} $ & $ 7.60 \times 10^{-1} $ & $ 1.44 \times 10^{-24} $ & $ 3.90 \times 10^{-24} $ & $ 1.47 \times 10^{2} $ & $ 2.78 \times 10^{2} $ & $ 1.9 \times 10^{15} $ & $ 0.34 $ & $201$ & $864$ \\
\texttt{EHP50.6} &  & $ 8.84 \times 10^{3} $ & $ 2.40 \times 10^{-2} $ & $ 5.06 \times 10^{1} $ & $ 2.53 \times 10^{0} $ & $ 1.26 \times 10^{-24} $ & $ 3.90 \times 10^{-24} $ & $ 6.77 \times 10^{2} $ & $ 3.14 \times 10^{2} $ &   & $ 0.90 $ & $201$ & $864$ \\
\texttt{EMP50.6} & $5$ & $ 8.84 \times 10^{3} $ & $ 2.40 \times 10^{-2} $ & $ 5.06 \times 10^{1} $ & $ 2.37 \times 10^{0} $ & $ 9.29 \times 10^{-25} $ & $ 3.90 \times 10^{-24} $ & $ 9.03 \times 10^{2} $ & $ 2.94 \times 10^{2} $ & $ 1.7 \times 10^{15} $ & $ 0.40 $ & $257$ & $864$ \\
\texttt{EMP175.0} & $5$ & $ 8.84 \times 10^{3} $ & $ 8.30 \times 10^{-2} $ & $ 1.75 \times 10^{2} $ & $ 8.70 \times 10^{0} $ & $ 1.18 \times 10^{-24} $ & $ 3.90 \times 10^{-24} $ & $ 9.16 \times 10^{2} $ & $ 3.12 \times 10^{2} $ & $ 2.4 \times 10^{8} $ & $ 0.31 $ & $201$ & $864$ \\
\texttt{EHP1750.5} &  & $ 8.84 \times 10^{3} $ & $ 8.30 \times 10^{-1} $ & $ 1.75 \times 10^{3} $ & $ 8.85 \times 10^{1} $ & $ 1.50 \times 10^{-24} $ & $ 3.90 \times 10^{-24} $ & $ 7.49 \times 10^{2} $ & $ 3.18 \times 10^{2} $ &   & $ 0.61 $ & $257$ & $864$ \\
\texttt{EMP1750.5} & $5$ & $ 8.84 \times 10^{3} $ & $ 8.30 \times 10^{-1} $ & $ 1.75 \times 10^{3} $ & $ 9.05 \times 10^{1} $ & $ 1.23 \times 10^{-24} $ & $ 3.90 \times 10^{-24} $ & $ 7.82 \times 10^{2} $ & $ 3.25 \times 10^{2} $ & $ 5.4 \times 10^{10} $ & $ 0.17 $ & $257$ & $864$ \\
\texttt{LMP1.0PV} & $5$ & $ 3.75 \times 10^{7} $ & $ 1.90 \times 10^{-5} $ & $ 9.90 \times 10^{-1} $ & $ 3.11 \times 10^{-3} $ & $ 4.83 \times 10^{-24} $ & $ 4.85 \times 10^{-27} $ & $ 4.30 \times 10^{2} $ & $ 1.97 \times 10^{1} $ & $ 2.8 \times 10^{15} $ & $ 0.48 $ & $201$ & $864$ \\
\texttt{LMP1.0} & $5$ & $ 3.75 \times 10^{7} $ & $ 1.90 \times 10^{-5} $ & $ 9.90 \times 10^{-1} $ & $ 3.94 \times 10^{-3} $ & $ 9.52 \times 10^{-24} $ & $ 4.85 \times 10^{-27} $ & $ 5.78 \times 10^{0} $ & $ 2.50 \times 10^{1} $ & $ 3.3 \times 10^{15} $ & $ 1.83 $ & $289$ & $1024$ \\
\texttt{LHP1.6} &  & $ 3.75 \times 10^{7} $ & $ 3.00 \times 10^{-5} $ & $ 1.56 \times 10^{0} $ & $ 2.93 \times 10^{-2} $ & $ 1.38 \times 10^{-24} $ & $ 4.85 \times 10^{-27} $ & $ 3.25 \times 10^{2} $ & $ 1.18 \times 10^{2} $ &   & $ 2.55 $ & $145$ & $320$ \\
\texttt{LMP2.0PV} & $5$ & $ 3.75 \times 10^{7} $ & $ 3.80 \times 10^{-5} $ & $ 1.98 \times 10^{0} $ & $ 1.14 \times 10^{-2} $ & $ 8.96 \times 10^{-25} $ & $ 4.85 \times 10^{-27} $ & $ 5.73 \times 10^{2} $ & $ 3.61 \times 10^{1} $ & $ 1.9 \times 10^{13} $ & $ 0.19 $ & $257$ & $864$ \\
\texttt{LMP2.0} & $5$ & $ 3.75 \times 10^{7} $ & $ 3.80 \times 10^{-5} $ & $ 1.98 \times 10^{0} $ & $ 1.50 \times 10^{-2} $ & $ 3.25 \times 10^{-24} $ & $ 4.85 \times 10^{-27} $ & $ 1.73 \times 10^{1} $ & $ 4.77 \times 10^{1} $ & $ 2.9 \times 10^{15} $ & $ 0.80 $ & $257$ & $864$ \\
\texttt{LHP3.4} &  & $ 3.75 \times 10^{7} $ & $ 6.45 \times 10^{-5} $ & $ 3.36 \times 10^{0} $ & $ 9.43 \times 10^{-2} $ & $ 2.19 \times 10^{-25} $ & $ 4.85 \times 10^{-27} $ & $ 2.22 \times 10^{2} $ & $ 1.76 \times 10^{2} $ &   & $ 0.64 $ & $145$ & $320$ \\
\texttt{LHP5.2} &  & $ 3.75 \times 10^{7} $ & $ 1.00 \times 10^{-4} $ & $ 5.21 \times 10^{0} $ & $ 1.36 \times 10^{-1} $ & $ 9.53 \times 10^{-26} $ & $ 4.85 \times 10^{-27} $ & $ 1.77 \times 10^{2} $ & $ 1.63 \times 10^{2} $ &   & $ 0.86 $ & $145$ & $320$ \\
\texttt{LHP3.4} &  & $ 3.75 \times 10^{6} $ & $ 1.40 \times 10^{-4} $ & $ 3.39 \times 10^{0} $ & $ 3.67 \times 10^{-2} $ & $ 2.19 \times 10^{-25} $ & $ 4.85 \times 10^{-27} $ & $ 1.47 \times 10^{2} $ & $ 6.81 \times 10^{1} $ &   & $ 3.83 $ & $145$ & $320$ \\
\texttt{LMP3.4} & $5$ & $ 3.75 \times 10^{6} $ & $ 1.40 \times 10^{-4} $ & $ 3.39 \times 10^{0} $ & $ 3.20 \times 10^{-2} $ & $ 2.11 \times 10^{-25} $ & $ 4.85 \times 10^{-27} $ & $ 3.35 \times 10^{1} $ & $ 5.94 \times 10^{1} $ & $ 1.1 \times 10^{16} $ & $ 2.31 $ & $241$ & $512$ \\
\texttt{LMP3.4PV} & $5$ & $ 3.75 \times 10^{6} $ & $ 1.40 \times 10^{-4} $ & $ 3.39 \times 10^{0} $ & $ 1.69 \times 10^{-2} $ & $ 3.11 \times 10^{-25} $ & $ 4.85 \times 10^{-27} $ & $ 1.28 \times 10^{2} $ & $ 3.13 \times 10^{1} $ & $ 2.4 \times 10^{15} $ & $ 0.23 $ & $257$ & $864$ \\
\texttt{LMP5.1} & $5$ & $ 3.75 \times 10^{6} $ & $ 2.10 \times 10^{-4} $ & $ 5.08 \times 10^{0} $ & $ 5.22 \times 10^{-2} $ & $ 1.04 \times 10^{-25} $ & $ 4.85 \times 10^{-27} $ & $ 1.05 \times 10^{2} $ & $ 6.46 \times 10^{1} $ & $ 2.1 \times 10^{15} $ & $ 1.30 $ & $201$ & $512$ \\
\texttt{LHP11.5} &  & $ 3.75 \times 10^{7} $ & $ 2.20 \times 10^{-4} $ & $ 1.15 \times 10^{1} $ & $ 3.25 \times 10^{-1} $ & $ 2.64 \times 10^{-26} $ & $ 4.85 \times 10^{-27} $ & $ 5.21 \times 10^{1} $ & $ 1.78 \times 10^{2} $ &   & $ 0.81 $ & $201$ & $512$ \\
\texttt{LMP6.0} & $5$ & $ 3.75 \times 10^{6} $ & $ 2.50 \times 10^{-4} $ & $ 6.05 \times 10^{0} $ & $ 7.92 \times 10^{-2} $ & $ 1.68 \times 10^{-25} $ & $ 4.85 \times 10^{-27} $ & $ 5.15 \times 10^{1} $ & $ 8.24 \times 10^{1} $ & $ 2.5 \times 10^{16} $ & $ 0.53 $ & $257$ & $864$ \\
\texttt{LMP8.0} & $5$ & $ 3.75 \times 10^{6} $ & $ 3.30 \times 10^{-4} $ & $ 7.98 \times 10^{0} $ & $ 1.10 \times 10^{-1} $ & $ 1.34 \times 10^{-25} $ & $ 4.85 \times 10^{-27} $ & $ 6.86 \times 10^{1} $ & $ 8.66 \times 10^{1} $ & $ 1.7 \times 10^{16} $ & $ 1.18 $ & $201$ & $512$ \\
\texttt{LMP8.9} & $5$ & $ 3.75 \times 10^{6} $ & $ 3.70 \times 10^{-4} $ & $ 8.95 \times 10^{0} $ & $ 1.24 \times 10^{-1} $ & $ 1.30 \times 10^{-25} $ & $ 4.85 \times 10^{-27} $ & $ 4.10 \times 10^{1} $ & $ 8.68 \times 10^{1} $ & $ 1.6 \times 10^{16} $ & $ 1.15 $ & $201$ & $512$ \\
\texttt{LHP9.4} &  & $ 3.75 \times 10^{6} $ & $ 3.90 \times 10^{-4} $ & $ 9.43 \times 10^{0} $ & $ 1.60 \times 10^{-1} $ & $ 2.60 \times 10^{-26} $ & $ 4.85 \times 10^{-27} $ & $ 1.00 \times 10^{2} $ & $ 1.07 \times 10^{2} $ &   & $ 2.87 $ & $201$ & $512$ \\
\texttt{LMP9.4PV} & $5$ & $ 3.75 \times 10^{6} $ & $ 3.90 \times 10^{-4} $ & $ 9.43 \times 10^{0} $ & $ 9.89 \times 10^{-2} $ & $ 3.18 \times 10^{-26} $ & $ 4.85 \times 10^{-27} $ & $ 9.87 \times 10^{1} $ & $ 6.59 \times 10^{1} $ & $ 2.8 \times 10^{15} $ & $ 0.22 $ & $201$ & $864$ \\
\texttt{LMP9.4} & $5$ & $ 3.75 \times 10^{6} $ & $ 3.90 \times 10^{-4} $ & $ 9.43 \times 10^{0} $ & $ 1.32 \times 10^{-1} $ & $ 1.12 \times 10^{-25} $ & $ 4.85 \times 10^{-27} $ & $ 7.33 \times 10^{1} $ & $ 8.81 \times 10^{1} $ & $ 1.6 \times 10^{11} $ & $ 0.93 $ & $201$ & $512$ \\
\texttt{LMP9.9} & $5$ & $ 3.75 \times 10^{6} $ & $ 4.10 \times 10^{-4} $ & $ 9.91 \times 10^{0} $ & $ 1.33 \times 10^{-1} $ & $ 1.18 \times 10^{-25} $ & $ 4.85 \times 10^{-27} $ & $ 2.94 \times 10^{1} $ & $ 8.43 \times 10^{1} $ & $ 1.5 \times 10^{16} $ & $ 4.28 $ & $201$ & $512$ \\
\texttt{LMP10.9} & $5$ & $ 3.75 \times 10^{6} $ & $ 4.50 \times 10^{-4} $ & $ 1.09 \times 10^{1} $ & $ 1.51 \times 10^{-1} $ & $ 9.63 \times 10^{-26} $ & $ 4.85 \times 10^{-27} $ & $ 4.68 \times 10^{1} $ & $ 8.71 \times 10^{1} $ & $ 1.5 \times 10^{16} $ & $ 3.50 $ & $201$ & $512$ \\
\texttt{LHP24.0} &  & $ 3.75 \times 10^{7} $ & $ 4.60 \times 10^{-4} $ & $ 2.40 \times 10^{1} $ & $ 4.43 \times 10^{-1} $ & $ 7.81 \times 10^{-27} $ & $ 4.85 \times 10^{-27} $ & $ 3.24 \times 10^{2} $ & $ 1.16 \times 10^{2} $ &   & $ 0.83 $ & $201$ & $512$ \\
\texttt{LMP12.1} & $5$ & $ 3.75 \times 10^{6} $ & $ 5.00 \times 10^{-4} $ & $ 1.21 \times 10^{1} $ & $ 1.65 \times 10^{-1} $ & $ 8.34 \times 10^{-26} $ & $ 4.85 \times 10^{-27} $ & $ 4.98 \times 10^{1} $ & $ 8.55 \times 10^{1} $ & $ 1.4 \times 10^{16} $ & $ 3.60 $ & $201$ & $512$ \\
\texttt{LMP15.0} & $5$ & $ 3.75 \times 10^{6} $ & $ 6.20 \times 10^{-4} $ & $ 1.50 \times 10^{1} $ & $ 2.00 \times 10^{-1} $ & $ 6.24 \times 10^{-26} $ & $ 4.85 \times 10^{-27} $ & $ 3.00 \times 10^{1} $ & $ 8.38 \times 10^{1} $ & $ 9.2 \times 10^{15} $ & $ 3.94 $ & $201$ & $512$ \\
\texttt{LMP17.9} & $5$ & $ 3.75 \times 10^{6} $ & $ 7.40 \times 10^{-4} $ & $ 1.79 \times 10^{1} $ & $ 2.33 \times 10^{-1} $ & $ 4.42 \times 10^{-26} $ & $ 4.85 \times 10^{-27} $ & $ 3.12 \times 10^{1} $ & $ 8.19 \times 10^{1} $ & $ 1.0 \times 10^{16} $ & $ 3.53 $ & $201$ & $512$ \\
\texttt{LHP9.4} &  & $ 3.75 \times 10^{5} $ & $ 8.40 \times 10^{-4} $ & $ 9.43 \times 10^{0} $ & $ 8.33 \times 10^{-2} $ & $ 1.59 \times 10^{-26} $ & $ 4.85 \times 10^{-27} $ & $ 1.25 \times 10^{2} $ & $ 5.55 \times 10^{1} $ &   & $ 1.78 $ & $145$ & $320$ \\
\texttt{LHP11.2} &  & $ 3.75 \times 10^{5} $ & $ 1.00 \times 10^{-3} $ & $ 1.12 \times 10^{1} $ & $ 1.17 \times 10^{-1} $ & $ 1.57 \times 10^{-26} $ & $ 4.85 \times 10^{-27} $ & $ 5.69 \times 10^{1} $ & $ 6.54 \times 10^{1} $ &   & $ 1.79 $ & $145$ & $320$ \\
\texttt{LHP24.2} &  & $ 3.75 \times 10^{6} $ & $ 1.00 \times 10^{-3} $ & $ 2.42 \times 10^{1} $ & $ 4.28 \times 10^{-1} $ & $ 7.50 \times 10^{-27} $ & $ 4.85 \times 10^{-27} $ & $ 2.54 \times 10^{2} $ & $ 1.11 \times 10^{2} $ &   & $ 1.52 $ & $145$ & $320$ \\
\texttt{LMP24.2} & $5$ & $ 3.75 \times 10^{6} $ & $ 1.00 \times 10^{-3} $ & $ 2.42 \times 10^{1} $ & $ 3.18 \times 10^{-1} $ & $ 6.05 \times 10^{-27} $ & $ 4.85 \times 10^{-27} $ & $ 1.17 \times 10^{2} $ & $ 8.27 \times 10^{1} $ & $ 2.9 \times 10^{15} $ & $ 0.77 $ & $257$ & $864$ \\
\texttt{LMP24.2PV} & $5$ & $ 3.75 \times 10^{6} $ & $ 1.00 \times 10^{-3} $ & $ 2.42 \times 10^{1} $ & $ 3.05 \times 10^{-1} $ & $ 5.54 \times 10^{-27} $ & $ 4.85 \times 10^{-27} $ & $ 1.54 \times 10^{2} $ & $ 7.91 \times 10^{1} $ & $ 3.3 \times 10^{15} $ & $ 0.21 $ & $257$ & $864$ \\
\texttt{LHP5.9} &  & $ 3.75 \times 10^{5} $ & $ 1.10 \times 10^{-3} $ & $ 5.94 \times 10^{0} $ & $ 3.78 \times 10^{-2} $ & $ 1.08 \times 10^{-25} $ & $ 4.85 \times 10^{-27} $ & $ 7.48 \times 10^{1} $ & $ 4.00 \times 10^{1} $ &   & $ 1.72 $ & $145$ & $320$ \\
\texttt{LHP50.8} &  & $ 3.75 \times 10^{6} $ & $ 2.10 \times 10^{-3} $ & $ 5.08 \times 10^{1} $ & $ 7.49 \times 10^{-1} $ & $ 3.86 \times 10^{-27} $ & $ 4.85 \times 10^{-27} $ & $ 2.20 \times 10^{2} $ & $ 9.27 \times 10^{1} $ &   & $ 0.72 $ & $257$ & $864$ \\
\texttt{LMP50.8} & $5$ & $ 3.75 \times 10^{6} $ & $ 2.10 \times 10^{-3} $ & $ 5.08 \times 10^{1} $ & $ 6.48 \times 10^{-1} $ & $ 2.69 \times 10^{-27} $ & $ 4.85 \times 10^{-27} $ & $ 1.66 \times 10^{2} $ & $ 8.02 \times 10^{1} $ & $ 6.7 \times 10^{10} $ & $ 0.38 $ & $257$ & $864$ \\
\texttt{LHP99.1} &  & $ 3.75 \times 10^{6} $ & $ 4.10 \times 10^{-3} $ & $ 9.91 \times 10^{1} $ & $ 1.44 \times 10^{0} $ & $ 3.34 \times 10^{-27} $ & $ 4.85 \times 10^{-27} $ & $ 2.84 \times 10^{2} $ & $ 9.15 \times 10^{1} $ &   & $ 2.19 $ & $201$ & $512$ \\
\texttt{LMP99.1} & $5$ & $ 3.75 \times 10^{6} $ & $ 4.10 \times 10^{-3} $ & $ 9.91 \times 10^{1} $ & $ 1.22 \times 10^{0} $ & $ 1.88 \times 10^{-27} $ & $ 4.85 \times 10^{-27} $ & $ 1.84 \times 10^{2} $ & $ 7.73 \times 10^{1} $ & $ 4.8 \times 10^{10} $ & $ 0.33 $ & $241$ & $864$ \\
\texttt{LHP991.4} &  & $ 3.75 \times 10^{6} $ & $ 4.10 \times 10^{-2} $ & $ 9.91 \times 10^{2} $ & $ 1.47 \times 10^{1} $ & $ 3.28 \times 10^{-27} $ & $ 4.85 \times 10^{-27} $ & $ 2.75 \times 10^{2} $ & $ 9.34 \times 10^{1} $ &   & $ 1.46 $ & $201$ & $512$ \\
\texttt{LMP991.4} & $5$ & $ 3.75 \times 10^{6} $ & $ 4.10 \times 10^{-2} $ & $ 9.91 \times 10^{2} $ & $ 1.20 \times 10^{1} $ & $ 1.68 \times 10^{-27} $ & $ 4.85 \times 10^{-27} $ & $ 2.09 \times 10^{2} $ & $ 7.59 \times 10^{1} $ & $ 1.5 \times 10^{10} $ & $ 0.48 $ & $241$ & $864$ \\
\bottomrule
  \end{tabular}
\end{table*}


\section{Mass multipole decomposition of the quadrupole formula}
\label{sec:quadrupole}


A popular way of computing the GW in 3D numerical simulations
(even if performed in spherical coordinates) is to use Cartesian components to
evaluate the quadrupole formula \citep[see e.g.][]{Oohara1997,Scheidegger2008, Scheidegger2010, Muller2012}\footnote{Note that Eq.~25 in \cite{Muller2012} contains some typos.}.
In this case, we can compute the two polarisations of the GW signal as \citep{Oohara1997}:
\begin{align}
h_{+}(T,{\bf X}) &= \frac{1}{R}\frac{G}{c^4} (\ddot Q_{\theta\theta} -\ddot Q_{\varphi\varphi}), \\
h_{\times}(T,{\bf X}) &= \frac{2}{R}\frac{G}{c^4} (\ddot Q_{\theta\varphi}),
\end{align}
while the necessary components of $Q_{ij}$ (expressed in an orthonormal spherical coordinate basis)
can be computed as a function of the Cartesian components \citep{Oohara1997}:
\begin{eqnarray}
Q_{\theta\theta} &=& (Q_{xx} \cos^2 \Phi + Q_{yy} \sin^2 \Phi + 2 Q_{xy} \sin\Phi \cos\Phi) \cos^2 \Theta \nonumber \\
&& + Q_{zz} \sin^2\Theta \nonumber \\
&& - 2 (Q_{xz} \cos \Phi + Q_{yz} \sin \Phi) \sin\Theta\cos\Theta, \\
Q_{\varphi\varphi} &=& Q_{xx} \sin^2 \Phi + Q_{yy} \cos^2 \Phi - 2 Q_{xy} \sin\Phi \cos\Phi,\\
Q_{\theta\varphi} &=& (Q_{yy}-Q_{xx}) \cos\Theta\sin\Phi\cos\Phi \nonumber \\
&&+ Q_{xy}\cos\Theta (\cos^2\Phi - \sin^2\Phi) \nonumber\\
&&+Q_{xz} \sin\Theta\sin\Phi - Q_{yz} \sin\Theta\cos\Phi.
\end{eqnarray}

The goal is to decompose $Q_{ij}$ in terms of the mass quadrupole moment $Q_{lm}$ given by Eq.\eqref{e:Qlm}.
For this purpose we write the $l=2$ spherical harmonics in term of $(x,y,z)$:
\begin{align}
Y_{20}(\theta,\varphi) &= \frac{3}{4}\sqrt{\frac{5}{\pi}} \frac{x^{zz}}{r^2},\\
Y_{21}(\theta,\varphi) &=-\frac{1}{2} \sqrt{\frac{15}{2\pi}} \frac{x^{xz}+i x^{yz}}{r^2}, \\
Y_{22} (\theta,\varphi) &= \frac{1}{4}\sqrt{\frac{15}{2\pi}} \frac{x^{xx}-x^{yy}+2 x^{xy}i}{r^2},\\
Y_{2-1} (\theta,\varphi)&= -Y_{21}^\star(\theta,\varphi), \\
Y_{2-2}(\theta,\varphi) &= Y_{22}^\star(\theta,\varphi),
\end{align}
where $x^{ij} \equiv  x^i x^j - 1/3\, r^2 \delta_{ij}$.
Solving for $x^{ij}$:
\begin{align}
x^{xx} &=  \frac{1}{3}\sqrt{\frac{\pi}{5}} \,r^2 \,\left [ \sqrt{6}(Y_{22}+Y_{2-2}) -2Y_{20})\right ]
,\\
x^{yy} &= - \frac{1}{3}\sqrt{\frac{\pi}{5}} \,r^2\, \left [ \sqrt{6}(Y_{22}+Y_{2-2}) +2Y_{20})\right ]
,\\
x^{zz} &=  \frac{4}{3} \sqrt{\frac{\pi}{5}} \,r^2\,Y_{20}
,\\
x^{xy} &= - \sqrt{\frac{2\pi}{15}} \,r^2\,(Y_{22} - Y_{2-2}) \,i
, \\
x^{xz} &= - \sqrt{\frac{2\pi}{15}} \,r^2\,(Y_{21} - Y_{2-1})
, \\
x^{yz} &=  \sqrt{\frac{2\pi}{15}} \,r^2\,(Y_{21} + Y_{2-1}) i\,
\end{align}
Using these relations the mass quadrupole can be expressed in terms of $Q_{lm}$ as
\begin{align}
Q_{xx} &= -\frac{2}{3} \sqrt{\frac{\pi}{5}}  (Q^R_{20} - \sqrt{6}\, Q^R_{22}),\\
Q_{yy} &= - \frac{2}{3} \sqrt{\frac{\pi}{5}} ( Q^R_{20} + \sqrt{6}\, Q^R_{22}),\\
Q_{zz} &= \frac{4}{3} \sqrt{\frac{\pi}{5}}   Q^R_{20},\\
Q_{xy} &= - 2 \sqrt{\frac{2\pi}{15}}   Q^I_{22}, \\
Q_{xz} &= -  2\sqrt{\frac{2\pi}{15}} \,Q^R_{21}, \\
Q_{yz} &=  2\sqrt{\frac{2\pi}{5}}  Q^I_{21},
\end{align}
where $Q_{lm}^{R}$ and $Q_{lm}^I$ are the real and imaginary part of $Q_{lm}$, respectively.
Finally, the strain from the two polarisations observed with
an angle $(\Theta,\Phi)$ is
\begin{align}
\begin{split}
h_+ &= \frac{1}{R}\frac{G}{c^4} \frac{1}{3}\sqrt{\frac{\pi}{5}} \bigg \{
6\, \ddot Q^R_{20} \sin^2\Theta  \\
&\quad + 2 \sqrt{6} \left[ \ddot Q^R_{21}\cos\Phi - \ddot Q^I_{21} \sin\Phi \right]\sin(2\Theta)  \\
&\quad+ \sqrt{6} \left[ \ddot Q^R_{22} \cos(2\Phi) - \ddot Q^I_{22} \sin(2\Phi)\right] \\
&\quad \left[3+\cos(2\Theta)\right] \bigg \}, \label{eq:hplus}
\end{split}\\
\begin{split}
h_{\times} &= - \frac{1}{R}\frac{G}{c^4} 4\sqrt{\frac{2\pi}{15}}  \bigg \{
  \left [ \ddot Q^R_{21} \sin\Phi  + \ddot Q^I_{21} \cos\Phi \right ] \sin\Theta  \\
&\quad  + \left[ \ddot Q^R_{22} \sin(2\Phi) + \ddot Q^I_{22} \cos(2\Phi) \right ] \cos\Theta \bigg \}. \label{eq:hcross}
\end{split}
\end{align}

Eqs.~(\ref{eq:hplus}-\ref{eq:hcross}) can be expressed in a very compact form in terms of the spin-weighted spherical harmonics with $s=-2$:
\begin{align}
h = h_+-i h_\times& =  \frac{1}{R} \frac{G}{c^4}\frac{8\pi}{5} \sqrt{\frac{2}{3}} \sum_{m=-2}^{+2} \ddot Q_{2m} \, {}^{}_{-2}Y^{2m}(\Theta,\Phi),
\end{align}
where we have used the values of the $l=2$ spin-weighted spherical harmonics with $s=-2$:
\begin{align}
{}^{}_{-2}Y^{20}(\theta,\varphi) &= \frac{1}{4}\sqrt{\frac{15}{2\pi}}\sin^2\theta, \\
{}^{}_{-2}Y^{2\pm1}(\theta,\varphi) &= \frac{1}{8}\sqrt{\frac{5}{\pi}}  (2 \sin\theta \pm \sin 2\theta) e^{\pm i\,\varphi}, \\
{}^{}_{-2}Y^{2\pm2}(\theta,\varphi) &= \frac{1}{16}\sqrt{\frac{5}{\pi}} (3 \pm 4 \cos\theta +\cos 2\theta) e^{\pm i\,2\varphi},
\end{align}
that fulfill the orthonormality relation
\begin{equation}
\int d\Omega \,\,{}^{}_sY^{lm} \,\,{}^{}_sY^{l'm'*} = \delta_{ll'} \delta_{mm'}.
\end{equation}
This decomposition naturally expresses $h$ in terms of the spin-weighted spherical harmonics with $s=-2$ (the only
ones capable of describing GWs) and $l=2$ (because our derivation comes from the quadrupole formula where
higher order $l$ are subdominant). This expression allows to connect our results with the usual expansion of $h$ in
spin-weighted spherical harmonics
\begin{equation}
    h ({\bf X}, T)= \sum_{l=2}^{\infty} h_{lm}(D, T)\,\, {}_{-2}Y^{lm} (\Theta,\Phi).
    \label{eq:hlm}
\end{equation}
The coefficients $h_{lm}$ (with $l=2$ in our case this case) can be directly related to the corresponding $Q_{2m}$,
and allow to study the impact of a perturbation with particular $m$ in the GW strain.

\bsp	
\label{lastpage}
\end{document}